\documentclass[11pt,a4paper]{article}
\pdfoutput=1
\usepackage[utf8]{inputenc}
\usepackage{jheppub}
\hypersetup{unicode=true,bookmarksopen=true}

\usepackage{cleveref}

\usepackage{amsthm}
\usepackage{mathtools}
\usepackage{lmodern}
\usepackage[T1]{fontenc}
\usepackage{bbm}
\DeclareMathAlphabet{\mathbfi}{OML}{cmm}{b}{it}

\usepackage[nointegrals]{wasysym}

\let\originalleft\left
\let\originalright\right
\renewcommand{\left}{\mathopen{}\mathclose\bgroup\originalleft}
\renewcommand{\right}{\aftergroup\egroup\originalright}
\makeatletter
\newcommand{\biggg}{\bBigg@\thr@@}
\newcommand{\Biggg}{\bBigg@{3.5}}
\makeatother

\makeatletter
\newenvironment{equations}[1][]{\subequations\ifx\relax#1\relax\else\label{#1}\fi\align\ignorespaces}{\endalign\ignorespacesafterend\endsubequations}
\def\@spliteq#1{\begin{equation}\begin{split}#1\end{split}\end{equation}}
\def\@spliteqstar#1{\begin{equation*}\begin{split}#1\end{split}\end{equation*}}
\def\splitequation{\collect@body\@spliteq}
\expandafter\def\csname splitequation*\endcsname{\collect@body\@spliteqstar}

\expandafter\def\csname endsplitequation*\endcsname{\ignorespacesafterend}
\makeatother

\renewcommand{\vec}[1]{{\ifnum9<1#1\mathbf{#1}\else\ifcat\noexpand#1\relax\boldsymbol{#1}\else\mathbfi{#1}\fi\fi}}
\newcommand{\mathe}{\mathrm{e}}
\newcommand{\mathi}{\mathrm{i}}
\let\oldre\Re
\let\oldim\Im
\renewcommand{\Re}{\mathop{\oldre\mathfrak{e}}}
\renewcommand{\Im}{\mathop{\oldim\mathfrak{m}}}
\newcommand{\total}{\mathop{}\!\mathrm{d}}
\newcommand{\abs}[1]{{\left\lvert{#1}\right\rvert}}
\newcommand{\norm}[1]{{\left\lVert{#1}\right\rVert}}
\newcommand{\1}{\mathbbm{1}}

\newcommand{\eqend}[1]{\,#1}

\newcommand{\bra}[1]{\left\langle{#1}\right\vert}
\newcommand{\ket}[1]{\left\vert{#1}\right\rangle}

\newcommand{\expect}[1]{\left\langle{#1}\right\rangle}
\newcommand{\normord}[1]{\mathopen{:}{#1}\mathclose{:}}
\newcommand{\supp}{\operatorname{supp}}

\newcommand{\esssup}{\operatornamewithlimits{ess\ sup}}
\newcommand{\defby}{\coloneq}
\newcommand{\bydef}{\eqcolon}

\makeatletter
\newcommand{\dhat}[1]{%
\begingroup%
  \let\macc@kerna\z@%
  \let\macc@kernb\z@%
  \let\macc@nucleus\@empty%
  \hat{\mathchoice%
    {\raisebox{.4ex}{\vphantom{\ensuremath{\displaystyle #1}}}}%
    {\raisebox{.4ex}{\vphantom{\ensuremath{\textstyle #1}}}}%
    {\raisebox{.32ex}{\vphantom{\ensuremath{\scriptstyle #1}}}}%
    {\raisebox{.28ex}{\vphantom{\ensuremath{\scriptscriptstyle #1}}}}%
    \smash{\hat{#1}}}%
\endgroup%
}
\makeatother

\makeatletter
\DeclareRobustCommand*{\citeDLMFeq}{\hyper@normalise\citeDLMFeq@}
\def\citeDLMFeq@#1#2{\cite[\hyper@linkurl{Eq.~#1.#2}{https://dlmf.nist.gov/#1.E#2}]{DLMF}}
\makeatother

\allowdisplaybreaks

\makeatletter
\gdef\@fpheader{\strut}
\makeatother

\makeatletter
\g@addto@macro\@floatboxreset\centering
\makeatother

\newtheorem{theorem}{Theorem}
\newtheorem{lemma}{Lemma}

\begin{document}

\title{Finite relative entropy for locally squeezed states}

\author[a]{Daniela Cadamuro,}
\author[a,b]{Markus B. Fr\"{o}b,}
\author[c]{Dimitrios Katsinis}
\author[d,1]{and Jan Mandrysch\note{Corresponding author.}}

\affiliation[a]{Institut f{\"u}r Theoretische Physik, Universit{\"a}t Leipzig, Br{\"u}derstra{\ss}e 16, 04103 Leipzig, Germany}
\affiliation[b]{Department Mathematik, Friedrich-Alexander-Universit{\"a}t Erlangen--N{\"u}rnberg, Cauerstra{\ss}e 11, 91058 Erlangen, Germany}
\affiliation[c]{National and Kapodistrian University of Athens, Department of Physics, 15784 Zografou, Attiki, Greece}
\affiliation[d]{Institute for Quantum Optics and Quantum Information, {\"O}sterreichische Akademie der Wissenschaften, Boltzmanngasse 3, 1090 Wien, Austria}

\emailAdd{cadamuro@itp.uni-leipzig.de}
\emailAdd{markus.froeb@fau.de}
\emailAdd{dkatsinis@phys.uoa.gr}
\emailAdd{jan.mandrysch@oeaw.ac.at}

\abstract{Relative entropy serves as a fundamental measure of state distinguishability in both quantum information theory and relativistic quantum field theory. Despite its conceptual importance, however, explicit computations of relative entropy remain notoriously difficult. Thus far, results in closed form have only been obtained for ground states, coherent states, and, more recently, single-mode squeezed states.

In this work, we extend the analysis to multi-mode squeezed states, imposing that the squeezing generators be local either in space or in spacetime, which results in a continuum of squeezed modes. We provide a detailed and self-contained analysis of such states for a free scalar quantum field on Minkowski spacetime, connecting also with older results on the essential self-adjointness of the Wick square, and showing that they lie in the folium of the Minkowski vacuum representation.

We uncover that local-in-space squeezing leads to physically unreasonable states which have a divergent energy density and relative entropy, while local-in-spacetime squeezing provides a physically reasonable class of states with finite relative entropy.}


\maketitle

\section{Introduction}

In this article we discuss squeezing from a foundational viewpoint and establish a well-behaved class of 'infinite-mode squeezed states' with finite and computable relative entropy. We prove our results for a real linear massive scalar field $\Phi$ in a Fock representation on Minkowski space in arbitrary dimension. We will focus on a class of states of the form
\begin{equation}\label{eq:deflocsqueeze}
W_2(h)\Omega \eqend{,} \quad W_2(h) \defby \exp\left[ \tfrac{\mathi}{2} \normord{\Phi^2}(h) \right] \eqend{,}
\end{equation}
where $\Omega$ denotes the Fock vacuum state, $\normord{\Phi^2}(h)$ denotes the Wick square smeared with a suitable Schwartz function $h \in \mathcal{S}_\mathbb{R}(\mathbb{R}^{D+1})$ in spacetime. As we will argue, such states may be produced by a local form of continuous or infinite-mode squeezing, which is why we refer to them as \emph{locally squeezed states}. It is also apparent that these states generalize the well-studied class of coherent states in QFT which are formed by acting with Weyl operators $W_1(h) = \exp\left[ \mathi \Phi(h) \right]$ on the Fock vacuum\footnote{A textbook account for this class of states, referred to as microscopic all-order coherent states, is \cite[Secs.~18.5.5, 26]{HR15}}.

While infinite-mode squeezing has been studied before \cite[and references therein]{VWW01,HR15}, also recently in the context of conformal field theory~\cite{hollands2019,Panebianco2020,HofmanVitouladitis2025,Vitouladitis2025}, surprisingly, \emph{locally} squeezed states seem to have received little attention so far.

A rigorous definition of locally squeezed states rests upon self-adjointness of the Wick square (to form the imaginary exponential that defines $W_2(h)$), where we adapt and generalize previous results in the literature \cite{BZ64,LS65,Kle73a,Bae89,San12} in order to show that locally squeezed states are quasi-free, live in the same Fock representation as $\Omega$ and can be produced by local Hamiltonians quadratic in $\Phi$.

We do this analysis first for states which are squeezed locally in \emph{space}. However, the resulting notion of squeezing changes the vacuum energy density by infinite amounts. Instead, we then look at states which are squeezed locally in \emph{spacetime}, as defined above. For those we establish that they have desirable properties of physically reasonable states by showing that they modify the vacuum correlations and its energy density smoothly. Moreover, for a Schwartz function $h$ supported in a wedge region $\mathcal{W}$, we show that both the total energy and the relative entropy of these states with respect to the Minkowski vacuum and the wedge algebra are finite.

The last result connects to a wider program on computing relative entropies in QFT. The relative entropy is the established generalization of von Neumann entropy to the setting of QFT and complies with fundamental principles like relativistic covariance and causality. However, it is well known to be notoriously challenging to compute and explicit results have so far been given only for ground, thermal and coherent states (see for example the recent Refs.~\cite{DFP18,longo2019,casinigrillopontello2019,hollands2019,hollandsishibashi2019,ciollilongoruzzi2019,Panebianco2020,DA21,KPV21a,ciollilongoranalloruzzi2022,bostelmanncadamurodelvecchio2022,galandamuchverch2023,garbarzpalau2023,froebmuchpapadopoulos2023,dangeloetal2024,brunettifredenhagenpinamonti2025} and references therein) as well as recently for single-mode squeezed states \cite{LLR21,GRSV25}. These all rely on restricting the state comparison to a simple type of spacetime region, where the relative entropy can be related to a geometric quantity like the energy density
\begin{equation}
\label{eq:energydensity}
T^{00}(x) = \frac{1}{2} \left[ (\partial_0 \Phi)^2 + \sum_{i=1}^D (\partial_i \Phi)^2 + m^2 \Phi^2 \right](x) \eqend{,}
\end{equation}
here given for a real scalar field $\Phi$ with mass $m > 0$. In particular, let $\mathcal{A}_{\mathcal{W}} \subset B(\mathcal{H})$ be the von Neumann algebra generated by Weyl operators $\exp\left[ \mathi \Phi(f) \right]$ for a wedge $\mathcal{W}$, i.e., with $\supp f \subset \mathcal{W}$; for concreteness we consider the canonical right wedge $\mathcal{W} = \{ x \colon x^1 > \abs{x^0} \}$. Now, let $U \in \mathcal{A}_\mathcal{W}$ be a unitary in the wedge algebra. Then the relative entropy between the state $U \Omega$ and the vacuum $\Omega$ restricted to the wedge algebra $\mathcal{A}_{\mathcal{W}}$ takes the form \cite{longo2019,casinigrillopontello2019}
\begin{equation}
\label{eq:relentropy}
S_\mathrm{rel}(U \Omega \Vert \Omega; \mathcal{A}_{\mathcal{W}}) = 2 \pi \bra{\Omega} U^\dagger \int_{x^0 = 0, x^1 \geq 0} x^1 \normord{T^{00}}(x) \total^D \vec{x} \, U \ket{\Omega} \eqend{,}
\end{equation}
where double dots around $T^{00}$ indicate normal ordering with respect to the vacuum state.

In contrast to von Neumann entropy which is UV-divergent in QFT, relative entropy remains finite in all cases studied in the references above although general results in this regard are missing. It is of course possible to introduce a UV cutoff which renders the von Neumann entropy finite, for example by working on a lattice instead of in the continuum. With such a cutoff in place, the entropy of squeezed states has been computed numerically~\cite{bianchihacklyokomizo2015,katsinispastrastetradis2023,katsinispastrastetradis2024}. However, the continuum limit remains divergent. In this work, we therefore focus on relative entropy and extend the previously mentioned results to states which are squeezed with respect to infinitely many modes, in particular, locally squeezed states. For this we establish that $W_2(h)$ for arbitrary $h \in \mathcal{S}_\mathbb{R}(\mathbb{R}^{D+1})$ is a unitary operator on the Fock space associated with $\Phi$ and moreover an element of $\mathcal{A}_{\mathcal{W}}$ if $h \in \mathcal{S}_\mathbb{R}(\mathcal{W})$. The relative entropy for the wedge algebra between a locally squeezed state and the vacuum can then be computed using \eqref{eq:relentropy} by setting $U = W_2(h)$.

The further outline of our article goes as follows. In \Cref{sec:qotoqft} we motivate the connection of locally squeezed states for the real scalar field $\Phi$ with squeezing as it is conventionally discussed in quantum optics and quantum information. We motivate here two variants of local squeezing --- either in space or in spacetime --- and discuss their properties separately in the two following sections. In \Cref{sec:cauchy}, we consider local squeezing in space and base our discussion on the initial value formalism for $\Phi$. In particular, we consider unitary squeezing operators associated with Hamiltonian densities which are local in space and quadratic in the time-zero fields. We show that the change in energy density produced by such an operation is infinite. Hence, also the total energy and relative entropy (in a wedge, with the squeezing supported in this wedge) will be infinite for this type of locally squeezed states, in contrast to the existing results for coherent states. In \Cref{sec:spacetime} we move on to our main results and discuss local squeezing in spacetime using the covariant formalism. We establish that local squeezing is unitarily implemented by the imaginary exponential of the (spacetime-)smeared Wick square and that it yields a well-defined class of quasi-free states. Moreover, we establish that local squeezing in spacetime produces smooth changes in the vacuum correlations and energy density as well as finite changes of the total energy and entropy restricted to the wedge. Finally, we summarize and discuss our results in \Cref{sec:conclusion}.

\section{Squeezing: From quantum optics to QFT}
\label{sec:qotoqft}

In quantum optics, \emph{squeezing} refers to the redistribution of uncertainties among quantum mechanical observables. In quantum information it is often considered a valuable resource since it may reduce uncertainty in selected observables below the ordinary Heisenberg limit; see, e.g., \cite[Ch.~7]{GK08}. A mathematical textbook account is \cite[Chs.~27--29]{HR15}.

It is typically discussed in the canonical formalism, where for a collection of (self-adjoint) quantum mechanical variables $(\phi,\pi) \defby (\phi_j,\pi_j)_{j=1}^N$ we assume canonical commutation relations (CCR):
\begin{equation}
[\phi_j,\phi_k]=[\pi_j,\pi_k] = 0 \eqend{,} \quad [\phi_j,\pi_k] = \mathi \delta_{jk} \eqend{,} \qquad j,k=1,\ldots,N \eqend{;}
\end{equation}
we call the resulting *-algebra $\mathrm{CCR}_N$. A \emph{squeezing transformation} with respect to $(\phi,\pi)$ is then any *-automorphism on $\mathrm{CCR}_N$ of the form
\begin{equation}
\phi_j \mapsto \mathe^{-r_j} \phi_j \eqend{,} \quad \pi_j \mapsto \mathe^{r_j} \pi_j \eqend{,} \quad r \defby (r_j)_{j=1}^N \subset \mathbb{R}^N \eqend{,}
\end{equation}
which implies a similar adjustment of the associated variances
\begin{equation}
\label{eq:mode_squeezing_variances}
\Delta \phi_j \mapsto \mathe^{-2r_j} \Delta \phi_j \eqend{,} \quad \Delta \pi_j \mapsto \mathe^{2r_j} \Delta \pi_j \eqend{,}
\end{equation}
which for a central observable ($\expect{A} = 0$) are given by $\Delta A \defby \expect{A^2}$.

In concrete representations of $\mathrm{CCR}_N$, these transformations may be implementable by the adjoint action of the squeezing operator
\begin{equation}
\label{eq:squeezingham}
S(r) = \exp\left[ \frac{\mathi}{2} \sum_{j=1}^N r_j \{ \phi_j, \pi_j \} \right] \eqend{,}
\end{equation}
where $\{\cdot,\cdot\}$ denotes the anticommutator. Operationally, the squeezing transformation may thus be viewed as the consequence of switching on and off a certain Hamiltonian which is quadratic in the canonical variables. More general squeezing operators arise when we apply squeezing as in \eqref{eq:squeezingham}, but with respect to a different set of canonical variables $( \tilde{\phi}, \tilde{\pi})$ (on the same CCR algebra). In this manner, we obtain the generalized squeezing operator
\begin{equation}
\label{eq:squeezinghammixed}
S(c) = \exp\left[ \frac{\mathi}{2} \sum_{j=1}^N c_{\phi,j} \phi_j^2 + c_{\pi,j} \pi_j^2 + c_{0,j} \{ \phi_j, \pi_j \} \right]
\end{equation}
for suitable real constants $c \defby (c_{\phi,j}, c_{\pi,j}, c_{0,j})_j$. Given a vector state $\ket{\Omega}$ or a density matrix state $\rho$, we can define the class of squeezed states with respect to $\ket{\Omega}$, respectively, $\rho$, by
\begin{equation}
    S(c) \ket{\Omega}, \qquad S(c) \rho S(c)^\dagger \eqend{.}
\end{equation}

Typically, squeezing is viewed as a non-relativistic quantum mechanical operation on the photon field and it is assumed (in approximation) that only a single ($N=1$) or a few ($N=2,3,4,\ldots$) modes of the photon field are relevant, while the rest is fully ignored. However, in this article we take a more foundational perspective. This means in particular two things to us. First, we will model the photon field as a quantum field which implies that we deal with \emph{infinitely many} modes ($N = \infty$). Second, while operations may be induced effectively from nonlocal Hamiltonians, only operations produced by local Hamiltonians are foundational. Therefore, we will focus on the class of squeezing operations which can be implemented by switching on and off a quadratic Hamiltonian which is \emph{local} in the fields. We distinguish here between locality in \emph{space} and in \emph{spacetime} and will discuss both cases separately in the following sections.

\section{Local squeezing in space}
\label{sec:cauchy}

In this section we discuss a real linear massive scalar field $\Phi$ on Minkowski space $\mathbb{M} \defby \mathbb{R}^{D+1}$, $D \in \mathbb{N}$, on a Cauchy surface, i.e., in the initial value or timeslice formalism. The formalism is based on a decomposition into canonical variables (infinitely many) and allows for a direct generalization of conventional squeezing as discussed in \Cref{sec:qotoqft}. However, as we will see, the resulting notion of infinite mode squeezing will be ill-defined so that we spare mathematical and computational details and postpone rigorous results to the next section where the covariant formalism is discussed.

Choosing a constant-time Cauchy surface $\Sigma = \mathbb{R}^D$ we may define canonical variables
\begin{equation}
\phi(\vec{x}) \defby \Phi(0,\vec{x}) \eqend{,} \quad \pi(\vec{x}) \defby \dot\Phi(0,\vec{x}) \eqend{,}
\end{equation}
which satisfy the canonical commutation relations
\begin{equation}
\label{eq:ccr}
[\phi(\vec{x}),\phi(\vec{y})] = [\pi(\vec{x}),\pi(\vec{y})] = 0 \eqend{,} \quad [\phi(\vec{x}),\pi(\vec{y})] = \mathi \delta^D(\vec{x}-\vec{y}) \eqend{,}
\end{equation}
which ensures that $(\phi,\pi) = (\phi(\vec{x}),\pi(\vec{x}))_\vec{x}$ generates the same *-algebra as the canonical field $(\Phi(t,\vec{x}))_{t,\vec{x}}$. Then, by analogy with \eqref{eq:squeezingham}, squeezing with respect to $(\phi,\pi)$ is implemented by the squeezing operator
\begin{equation}
\label{eq:timeslicesqueezing}
S(r) = \exp\left[ \frac{\mathi}{2} \int r(\vec{x}) \{ \phi(\vec{x}), \pi(\vec{x}) \} \total^D \vec{x} \right] \eqend{,} \quad r \colon \mathbb{R}^D \to \mathbb{R} \eqend{,}
\end{equation}
whose adjoint action results in the transformed fields
\begin{equation}
\phi_S(\vec{x}) \defby S(r)^\dagger \phi(\vec{x}) S(r) = \mathe^{-r(\vec{x})} \phi(\vec{x}) \eqend{,} \quad \pi_S(\vec{x}) \defby S(r)^\dagger \pi(\vec{x}) S(r) = \mathe^{r(\vec{x})} \pi(\vec{x}) \eqend{.}
\end{equation}
Smearing with a Schwartz function $f \in \mathcal{S}_\mathbb{R}(\mathbb{R}^D)$ according to
\begin{equation}
\phi(f) \defby \int \phi(\vec{x}) f(\vec{x}) \total^D \vec{x} \eqend{,} \quad \pi(f) \defby \int \pi(\vec{x}) f(\vec{x}) \total^D \vec{x} \eqend{,}
\end{equation}
we therefore have
\begin{equation}
\phi_S(f) = \phi(f_{-r}) \eqend{,} \quad \pi_S(f) = \pi(f_r) \eqend{,} \quad f_r(\vec{x}) \defby \mathe^{r(\vec{x})} f(\vec{x}) \eqend{,}
\end{equation}
and compute the transformed variances
\begin{equation}
\Delta(\phi_S(f)) = \Delta(\phi(f_{-r})) \eqend{,} \quad \Delta(\pi_S(f)) = \Delta(\pi(f_r)) \eqend{,}
\end{equation}
where $\Delta(A) \defby \bra{ \Omega} A^2 \ket{\Omega } - \left( \bra{\Omega} A \ket{\Omega} \right)^2$.

The smearing is required to give finite results since vacuum expectation values of pointwise squares such as $\bra{\Omega} \phi^2(\vec{x}) \ket{\Omega}$ are generally divergent. Well-defined local squares are obtained by normal ordering which, in `unsmeared' notation, can be defined as
\begin{equation}
\normord{ A B }(\vec{x}) = \lim_{\vec{y} \to \vec{x}} \normord{ A(\vec{x}) B(\vec{y}) } \eqend{,} \quad \normord{ A(\vec{x}) B(\vec{y}) } \defby A(\vec{x}) B(\vec{y}) - \bra{\Omega} A(\vec{x}) B(\vec{y}) \ket{\Omega} \1 \eqend{.}
\end{equation}
Note that the integrand in \eqref{eq:timeslicesqueezing} is normally ordered as it stands, because in the standard Minkowski vacuum state $\ket{\Omega}$ it holds that $\bra{\Omega} \{ \phi(\vec{x}), \pi(\vec{y}) \} \ket{\Omega} = 0$.

In analogy with \eqref{eq:squeezinghammixed}, for squeezing with respect to a different set of canonical variables (for example taking a different Cauchy surface) we obtain a squeezing operator of the form
\begin{equation}
S(k) \defby \exp\left( \mathi \overline{A(k)} \right) \eqend{,} \quad A(k) \defby \frac{1}{2} \int \left( k_\phi(\vec{x}) \normord{\phi^2}(\vec{x}) + k_\pi(\vec{x}) \normord{\pi^2}(\vec{x}) + k_0(\vec{x}) \{ \phi, \pi \}(\vec{x}) \right) \total^D \vec{x}
\end{equation}
with real coefficient functions $k = ( k_\phi, k_\pi, k_0 )$, each mapping $\mathbb{R}^D \to \mathbb{R}$. We note here that, in view of well-known results on (essential) self-adjointness, in particular \cite[Sec.~5]{Kle73a}, $S(k)$ is unambiguously defined through spectral calculus in terms of the closure of $A(k)$.

With this ansatz for the squeezing operator we define our \emph{locally squeezed fields} by
\begin{equation}
\label{eq:locsqfields}
\phi_S(\vec{x}) \defby S(k)^\dagger \phi(\vec{x}) S(k) \eqend{,} \quad \pi_S(\vec{x}) \defby S(k)^\dagger \pi(\vec{x}) S(k) \eqend{.}
\end{equation}
A straightforward computation given in \Cref{app:sqspace}, based on the canonical commutation relations \eqref{eq:ccr} and $[ \normord{A^2}(\vec{x}), B ] = \lim_{\vec{y} \to \vec{x}} [ A(\vec{x}) A(\vec{y}), B ]$, yields
\begin{equations}
\phi_S(\vec{x}) &= k_{c,-}(\vec{x}) \phi(\vec{x}) - k_{s,\pi}(\vec{x}) \pi(\vec{x}) \eqend{,} \\
\pi_S(\vec{x}) &= k_{c,+}(\vec{x}) \pi(\vec{x}) + k_{s,\phi}(\vec{x}) \phi(\vec{x}) \eqend{,}
\end{equations}
where we defined the combinations
\begin{equations}
k_{c,\pm}(\vec{x}) &= \cosh \sqrt{\kappa(k,\vec{x})} \pm k_0(\vec{x}) \kappa(k,\vec{x})^{-\frac{1}{2}} \sinh \sqrt{\kappa(k,\vec{x})} \eqend{,} \\
k_{s,\phi}(\vec{x}) &= k_\phi(\vec{x}) \kappa(k,\vec{x})^{-\frac{1}{2}} \sinh \sqrt{\kappa(k,\vec{x})} \eqend{,} \\
k_{s,\pi}(\vec{x}) &= k_\pi(\vec{x}) \kappa(k,\vec{x})^{-\frac{1}{2}} \sinh \sqrt{\kappa(k,\vec{x})}  \eqend{,}
\end{equations}
as well as
\begin{equation}
\kappa(k,\vec{x}) \defby k_0(\vec{x})^2 - k_\phi(\vec{x}) k_\pi(\vec{x}) \eqend{.}
\end{equation}
If $k_0$, $k_\phi$ and $k_\pi$ are real functions, also $k_{c,\pm}$, $k_{s,\phi}$ and $k_{s,\pi}$ are real functions. Note also that $k_{c,\pm}$, $k_{s,\phi}$ and $k_{s,\pi}$ depend analytically on $k_0$, $k_\phi$ and $k_\pi$ since $\cosh \sqrt{\kappa}$ and $\kappa^{-1/2} \sinh \sqrt{\kappa}$ depend analytically on $\kappa$. It follows straightforwardly that the locally squeezed fields $(\phi_S,\pi_S)$ obey the canonical commutation relations \eqref{eq:ccr} iff $(\phi,\pi)$ do, and are thus local fields. In particular, they are in the Borchers class of the free field~\cite{borchers1960,epstein1963}.

We also consider the squeezing of normal-ordered products of the fields and, in particular, determine the squeezed energy density
\begin{equation}
T^{00}_S(0,\vec{x}) \defby S(k)^\dagger \normord{ T^{00}(0,\vec{x}) } S(k) \eqend{.}
\end{equation}
However, as shown in Appendix~\ref{app:sqspace} its vacuum expectation value is divergent,
\begin{equation}
\bra{\Omega} T_S^{00}(0,\vec{x}) \ket{\Omega} = \infty \eqend{,}
\end{equation}
unless $k$ is such that $S(k) \Omega = \mathe^{\mathi c} \Omega$ for some $c \in \mathbb{R}$, i.e., \emph{no squeezing at all}.

The class of states squeezed locally in space is therefore to be dismissed as physically nonreasonable (see e.g. \cite{BDFY15}). Moreover, the entropy of a state squeezed locally in space relative to the Minkowski vacuum is expected to be generally infinite as indicated by the special case where $k$ is supported in the right half-space. In this case, $S(k)$ acts within the algebra of the canonical right wedge $\mathcal{W} = \{ x \in \mathbb{R}^{D+1} \colon x^1 > \abs{x^0} \}$ (which for initial data at $x^0 = 0$ is simply the region $x^1 > 0$), and thus
\begin{equation}
S_\text{rel}\left( S(k) \Omega \Vert \Omega \right) = 2 \pi \int_{x^1 > 0} x^1 \bra{\Omega} T_S^{00}(0,\vec{x}) \ket{\Omega} \total^D \vec{x} = \infty \eqend{,}
\end{equation}
unless again $k$ is such that the squeezing operator $S(k)$ becomes a trivial phase shift.

\section{Local squeezing in spacetime}
\label{sec:spacetime}

We now consider a real linear scalar field $\Phi$ of mass $m$ on Minkowski spacetime $\mathbb{R}^d$ in $d = D+1$ dimensions (with $D \geq 1$) in the covariant formalism. We assume $\Phi$ to be in the standard GNS Fock representation whose (cyclic and separating) ground state vector is the unique Lorentz-invariant Minkowski vacuum state $\Omega$. This means in particular that for all real-valued Schwartz functions $f \in \mathcal{S}_\mathbb{R}(\mathbb{R}^d)$ the operator $\Phi(f)$ is an essentially self-adjoint operator on Fock space $\mathcal{F}$, which depends (real-)linearly on $f$. The Minkowski vacuum state $\omega$ is quasi-free and therefore fully specified by its two-point function $\omega_2$, which takes the form\footnote{Here and in the following, we denote the Minkowski inner product as $kx \defby - k^0 x^0 + \vec{k} \cdot \vec{x}$. For Schwartz functions $f, g$ on $\mathbb{R}^d$, we use the conventions:
\begin{equation*}
(f,g) \defby \int \overline{f(x)} g(x) \total^d x \eqend{,} \quad \tilde{f}(k) \defby \int f(x) \, \mathe^{- \mathi k x} \total^d x \eqend{,} \quad (f \ast g) (x) \defby \int f(x-y) g(y) \total^d y = \int \tilde{f}(p) \tilde{g}(p) \mathe^{\mathi p x} \frac{\total^d p}{(2\pi)^d} \eqend{,}
\end{equation*}
and we use the same notation also if $f$ or $g$ are tempered distributions, where the scalar product is defined by the application of the distribution and the Fourier transform by duality.}
\begin{equation}
\label{eq:2ptfunc}
\omega_2(f,g) \defby \bra{\Omega} \Phi(f) \Phi(g) \ket{\Omega} = \int \tilde{f}(-\omega_\vec{p},-\vec{p}) \tilde{g}(\omega_\vec{p},\vec{p}) \total \mu_m(\vec{p})
\end{equation}
with
\begin{equation}
\total \mu_m(\vec{p}) \defby \frac{1}{2 \omega_\vec{p}} \frac{\total^D \vec{p}}{(2\pi)^D} \eqend{,} \quad \omega_\vec{p} \defby \sqrt{ \vec{p}^2 + m^2 } \eqend{.}
\end{equation}

For real $f$ and $g$, the Pauli--Jordan commutator function $\Delta$ can be obtained from the imaginary part of the two-point function according to
\begin{equation}
\label{eq:delta_def}
\Delta(f,g) \defby - \mathi \bra{\Omega} \bigl[ \Phi(f), \Phi(g) \bigr] \ket{\Omega} = 2 \Im \omega_2(f,g) \eqend{.}
\end{equation}
The space $(\mathcal{S}_\mathbb{R}(\mathbb{R}^d),\Delta)$ is then a presymplectic space, the phase space associated with $\Phi$, which becomes a symplectic space by dividing out the kernel of $\Delta$. Completing this space with respect to the norm $\norm{\cdot}_\omega$ induced by $\omega_2$, i.e. $\norm{ f }_\omega \defby \sqrt{ \omega_2(f,f) }$, gives rise to the one-particle Hilbert space
\begin{equation}
\label{eq:1partcompl}
\mathcal{H} = \overline{\mathcal{S}_\mathbb{R}(\mathbb{R}^d)/\ker \Delta}^{\norm{\cdot}_\omega} \eqend{.}
\end{equation}
By definition, this implies that $\mathcal{S}_\mathbb{R}(\mathbb{R}^d)$ maps onto a dense subset of $\mathcal{H}$. The Hilbert space $\mathcal{F}$ of the free scalar QFT is then obtained as the usual symmetric Fock space over $\mathcal{H}$.

By the Schwartz kernel theorem, $\omega_2$ and $\Delta$ have unique distributional integral kernels (which we denote by the same symbols) such that $\omega_2(f,g) = (f,\omega_2 \ast g)$ and $\Delta(f,g) = (f, \Delta \ast g)$. In particular, we can define $\Delta$ in terms of the Fourier transform of this integral kernel, given by
\begin{equation}
\label{eq:Delta}
\tilde{\Delta}(k) = - \mathi \frac{\pi}{\omega_\vec{k}} \left[ \delta\left( - \omega_\vec{k} + k^0 \right) - \delta\left( \omega_\vec{k} + k^0 \right) \right] \eqend{.}
\end{equation}
From this explicit expression, we see immediately that the kernel of $\Delta$ consists of all those Schwartz functions whose Fourier transform at $k^0 = \pm \omega_\vec{k}$ vanishes:
\begin{equation}
\ker \Delta = \left\{ f \in \mathcal{S}_{\mathbb{R}}(\mathbb{R}^d) \colon \tilde{f}(\omega_\vec{k},\vec{k}) = \tilde{f}(-\omega_\vec{k},\vec{k}) = 0 \right\} \eqend{.}
\end{equation}
We will also need the the symmetric bilinear form
\begin{equation}
\label{eq:mu_def}
\mu(f,g) = \int f(x) \mu(x-y) g(y) \total^d x \total^d y \eqend{,}
\end{equation}
which we also specify by writing the Fourier transform of its integral kernel
\begin{equation}
\tilde{\mu}(k) = \frac{\pi}{2\omega_{\vec{k}}} \left[ \delta\left( - \omega_\vec{k} + k^0 \right) + \delta\left( \omega_\vec{k} + k^0 \right) \right] \eqend{.}
\end{equation}

We begin with the commutation relations between the field and the Wick square. We recall the definition of the Wick square,
\begin{equation}
\normord{\Phi^2}(h) \defby \lim_{f\otimes f' \to h_2} \, \left[ \Phi(f) \Phi(f') - \left( \Omega, \Phi(f) \Phi(f') \Omega \right) \1 \right] \eqend{,}
\end{equation}
where $h_2(x,y) \defby h(x) \delta(x-y)$. It is then straightforward to infer the commutator relation
\begin{splitequation}
\label{eq:wickcomm}
\left[ \tfrac{1}{2} \normord{\Phi^2}(h), \Phi(g) \right] &= \frac{1}{2} \lim_{f \otimes f' \to h_2} \left[ \Phi(f) \Phi(f'), \Phi(g) \right] \\
&= \frac{\mathi}{2} \lim_{f \otimes f' \to h_2} \left[ (f, \Delta \ast g) \Phi(f') + (f',\Delta \ast g) \Phi(f) \right] \\
&= \mathi \Phi(t_h g)
\end{splitequation}
with the real-linear operator $t_h \colon \mathcal{S}_\mathbb{R}(\mathbb{R}^d) \to \mathcal{S}_\mathbb{R}(\mathbb{R}^d)$ given by
\begin{equation}
\label{eq:th1}
(t_h g)(x) \defby h(x) (\Delta \ast g)(x) = h(x) \int \Delta(x-y) g(y) \total^d y \eqend{.}
\end{equation}
Analogously, for the iterated commutators defined in Eq.~\eqref{eq:itcomm} we find
\begin{equation}
\label{eq:wickitcomm}
\left[ \tfrac{1}{2} \normord{\Phi^2}(h), \Phi(g) \right]_n = \mathi^n \Phi(t_h^n g) \eqend{.}
\end{equation}

Since $\Phi(g)$ and $\normord{\Phi^2}(h)$ are unbounded operators, the commutation relations in \eqref{eq:wickcomm} and \eqref{eq:wickitcomm} can only hold on certain domains, in particular, on any common invariant domain $\mathcal{D} \subset \mathcal{F}$. Based on well-known results on (essential) self-adjointness of the field $\Phi$ and the Wick square, which we mentioned in the introduction and state more explicitly in \Cref{lem:wicksadj} below, we can use the finite-particle subspace $\mathcal{F}_0 \subset \mathcal{F}$ for $\mathcal{D}$, which is dense in $\mathcal{F}$~\cite[Thm.~X.41]{reedsimon2}, \cite[Prop.~5.2.2]{bratellirobinson2}.

Multiplying~\eqref{eq:wickitcomm} by $1/n!$ and formally summing over $n$, we can phrase the commutator in terms of a formal power series in $s$ as
\begin{equation}
\label{eq:bch}
\exp\left[ - \frac{\mathi s}{2} \normord{\Phi^2}(h) \right] \Phi(g) \exp\left[ \frac{\mathi s}{2} \normord{\Phi^2}(h) \right] = \Phi\left( T_{sh} g \right) \eqend{,}
\end{equation}
where we defined (formally)
\begin{equation}
\label{eq:th2}
T_h \defby \sum_{n=0}^\infty \frac{1}{n!} t_h^n = \exp(t_h) \eqend{.}
\end{equation}
Performing a further formal sum, we obtain the adjoint action on the exponentials of the field, at least as a formal power series in $r$ and $s$:
\begin{equation}
\label{eq:bch2}
\exp\left[ - \frac{\mathi s}{2} \normord{\Phi^2}(h) \right] \exp\left[ \mathi r \Phi(g) \right] \exp\left[ \frac{\mathi s}{2} \normord{\Phi^2}(h) \right] = \exp \left[ \mathi r \Phi\left( T_{sh} g \right) \right] \eqend{.}
\end{equation}
We think of $T_h$ as representing local squeezing (in spacetime) as a transformation of the underlying phase space $(\mathcal{S}_\mathbb{R}/\ker \Delta,\Delta)$. In \Cref{sec:bog}, we establish that $T_h$ is indeed a well-defined phase space transformation which induces Bogoliubov transformations on the field algebra. This implies that we can define locally squeezed states abstractly. Then, in \Cref{sec:uimpl}, we establish that \eqref{eq:bch} and \eqref{eq:bch2} hold for sufficiently small $r$ and $s$ (depending on $h$) as operator equations on $\mathcal{F}$, which implies that local squeezing can be unitarily implemented. In particular, we can represent locally squeezed states as vector states $\exp\left( \tfrac{\mathi}{2} \normord{\Phi^2}(h) \right) \Omega \in \mathcal{F}$. In \Cref{sec:finiteentropy}, as our main result, we show that local squeezing produces smooth changes in the vacuum correlations and the vacuum energy density as well as finite changes of the energy and the entropy, in the latter case restricted to the wedge.

\subsection{Local squeezing as a Bogoliubov transformation}
\label{sec:bog}

\begin{lemma}
\label{lem:thdef}
For any $h \in \mathcal{S}_\mathbb{R}(\mathbb{R}^d)$, the operators $t_h$ defined by \eqref{eq:th1} and $T_h$ defined by \eqref{eq:th2} are bounded linear operators on the single-particle Hilbert space $\mathcal{H}$ defined after Eq.~\eqref{eq:1partcompl}.
\end{lemma}
\begin{proof}
We recall that the norm on $\mathcal{H}$ is given by
\begin{equation}
\norm{ f }_\omega^2 = \int \frac{1}{2 \omega_\vec{p}} \abs{ \tilde{f}(\omega_\vec{p},\vec{p}) }^2 \frac{\total^D \vec{p}}{(2 \pi)^D} \eqend{.}
\end{equation}
In \Cref{app:normestimates}, we compute the estimate~\eqref{eq:thf_norm_2_estimate}
\begin{equation}
\norm{ t_h f }_\omega \leq c(h) \norm{ f }_\omega
\end{equation}
with the constant~\eqref{eq:ch_def}
\begin{equation}
c(h) = \sup_{\vec{p},\vec{k}} \left( \frac{\omega_\vec{p}^{D+1}}{m^2} \abs{ \tilde{h}(\pm \omega_{\vec{p}+\vec{k}} \pm \omega_{\vec{k}}, \vec{p}) } \right) < \infty \eqend{,}
\end{equation}
which shows that $t_h$ is bounded on $\mathcal{H}$ with operator norm $\norm{ t_h } \leq c(h)$.  Furthermore, it follows that
\begin{equation}
\norm{ T_h f }_\omega \leq \sum_{k=0}^\infty \frac{1}{k!} \norm{ t_h^k f }_\omega \leq \sum_{k=0}^\infty \frac{c(h)^k}{k!} \norm{f}_\omega = \mathe^{c(h)} \norm{f}_\omega < \infty \eqend{,}
\end{equation}
such that also $T_h$ is a bounded operator on $\mathcal{H}$ with operator norm $\norm{ T_h } \leq \exp\left[ c(h) \right]$.
\end{proof}

It then easily follows that $T_h$ is a proper phase space transformation:
\begin{lemma}
\label{lem:sympl}
$T_h$ is a (bounded) linear symplectic transformation of $(\mathcal{S}_{\mathbb{R}}/\ker \Delta, \Delta)$.
\end{lemma}
\begin{proof}
By \Cref{lem:thdef}, $t_h$ and $T_h$ are bounded on $\mathcal{S}_\mathbb{R}$. That $t_h$~\eqref{eq:th1} is skew-symmetric with respect to $\Delta$ follows by construction:
\begin{splitequation}
\Delta(t_h f, g) &= \left( t_h f, \Delta \ast g \right) = \int h(x) ( \Delta \ast f )(x) ( \Delta \ast g )(x) \total^d x \\
&= \left( \Delta \ast f, t_h g \right) = - \left( f, \Delta \ast ( t_h g ) \right) = - \Delta(f, t_h g) \eqend{,}
\end{splitequation}
where we used that $\Delta(f,g) = \left( f, \Delta \ast g \right) = - \left( \Delta \ast f, g \right) = - \Delta(g,f)$ for real-valued $f$ and $g$. This implies that $T_h$ is symplectic with respect to $\Delta$:
\begin{splitequation}
\Delta(T_h f, g) = \Delta(\exp (t_h) f, g) = \Delta(f, \exp(-t_h) g) = \Delta(f, T_h^{-1}g) \eqend{.}
\end{splitequation}
\end{proof}
It follows that $T_h$ induces Bogoliubov transformations. We define the algebra of observables as the Weyl algebra
\begin{equation}
\label{eq:algdef}
\mathcal{A} \defby \left\{ W_1(f) = \exp\left( \mathi \overline{\Phi(f)} \right) \colon f \in \mathcal{S}_\mathbb{R}(\mathbb{R}^d) \right\}'' \subset B(\mathcal{F}) \eqend{,}
\end{equation}
where the precise definition of the Weyl operators $W_1$ is given in the next section. Moreover, we define the Bogoliubov transformations induced by $T_h$ through
\begin{equation}
\label{eq:bogoliubov_w1}
\alpha_{T_h} \left( W_1(f) \right) \defby W_1(T_h f) \eqend{.}
\end{equation}

\begin{theorem}
\label{thm:bogtrafo}
Let $\Phi$ be a real scalar massive quantum field in the Minkowski Fock vacuum representation in dimension $d$. For any real-valued Schwartz function $h \in \mathcal{S}_{\mathbb{R}}(\mathbb{R}^d)$, the transformation $T_h$ defined in \eqref{eq:th2} induces an automorphism $\alpha_{T_h}$ on the Weyl algebra $\mathcal{A}$ defined in~\eqref{eq:algdef}. The transformed Weyl operators defined by
\begin{equation}
W_{1,T_h}(f) = \alpha_{T_h}(W_1(f)) \defby W_1(T_h f)
\end{equation}
generate the same algebra, and the transformed state defined by
\begin{equation}
\omega_{T_h} \defby \omega \circ \alpha_{T_h} \eqend{,} \qquad \omega \defby \bra{\Omega} \cdot \ket{\Omega}
\end{equation}
is also quasi-free with two-point function
\begin{equation}
\omega_{2,T_h}(f, g) = \omega_2( T_h f, T_h g ) \eqend{.}
\end{equation}
\end{theorem}
\begin{proof}
Based on \Cref{lem:thdef} and \Cref{lem:sympl}, most parts of the proof are a standard exercise in the theory of Bogoliubov transformations. In particular, any linear symplectic transformation $T$ of a symplectic space $X$ induces an algebra automorphism $\alpha_T$ on the Weyl algebra $\mathcal{A}_{\{X\}}$ over $X$ \cite[Secs.~18.1--2]{HR15}, and we simply take $T = T_h$, $X = (\mathcal{S}_{\mathbb{R}}/\ker \Delta, \Delta)$ and $\mathcal{A}_{\{X\}} = \mathcal{A}$. That the transformed state $\omega_{T_h}$ is a state and quasi-free is a direct consequence of the identities
\begin{equations}
\omega_{T_h}\left( W_1(f) \right) &= \omega\left( W_1(T_h f) \right) = \mathe^{- \frac{1}{2} \omega_2(T_h f, T_h f)} = \mathe^{- \frac{1}{2} \omega_{2,T_h}(f, f)} \eqend{,} \\
\omega_{2,T_h}(f,f) \, \omega_{2,T_h}(g,g) &= \omega_2(T_h f, T_h f) \, \omega_2(T_h g, T_h g) \geq \frac{1}{4} \left[ \Delta( T_h f, T_h g ) \right]^2 = \frac{1}{4} \left[ \Delta(f,g) \right]^2 \eqend{,}
\end{equations}
where the second relation in both equations holds since $\omega$ is a quasi-free state with two-point function $\omega_2$, and the third relation in the second equation holds because $T_h$ is symplectic with respect to $\Delta$ by \Cref{lem:sympl}.
\end{proof}

To compute the relative entropy associated with the transformed states $\omega_{T_h}$, we will also need some information on the localization of $\alpha_{T_h}$. For a wedge region $\mathcal{W} \subset \mathbb{R}^{D+1}$, let $\mathcal{A}_\mathcal{W} \subset \mathcal{A}$ denote the associated wedge algebra defined as in \eqref{eq:algdef}, but restricted to functions $f$ with $\supp f \subset \mathcal{W}$, which we denote by $f \in \mathcal{S}_\mathbb{R}(\mathcal{W})$. From the definitions~\eqref{eq:th1} and~\eqref{eq:th2} of $t_h$ and $T_h$, it is then clear that if $f,h \in \mathcal{S}_\mathbb{R}(\mathcal{W})$, also $\supp t_h f, \supp T_h f \subset \mathcal{W}$. It follows that $\alpha_{T_h}$ preserves the wedge algebra $\mathcal{A}_\mathcal{W}$ for such $h$.

\subsection{Unitary implementability of local squeezing}
\label{sec:uimpl}

We now would like to show that the Bogoliubov transformation $\alpha_{T_h}$ given by \eqref{eq:bogoliubov_w1} can be unitarily implemented, i.e., that \eqref{eq:bch} and \eqref{eq:bch2} hold with a well-defined unitary operator $W_2(h)$ formally given by $\mathe^{\frac{\mathi}{2} \normord{ \Phi^2 }(h)}$. For this we make use of well-known results on (essential) self-adjointness of the field and the Wick square on Minkowski space, in particular \cite[App.~1]{LS65}. However, since Ref.~\cite{LS65} makes some unnecessary (partially implicit) assumptions, we generalize their arguments in \Cref{app:wickestimates}, and obtain
\begin{lemma}
\label{lem:wicksadj}
Let $\mathcal{F}_0 \subset \mathcal{F}$ denote the finite-particle subspace of the Fock space. Consider $f_j, h_j \in \mathcal{S}_\mathbb{R}(\mathbb{R}^d)$ for $j=1,\ldots,k$ and a vector $\psi \in \mathcal{F}_0$ containing at most $n$ particles. Then we have the bound
\begin{splitequation}
\label{eq:wicksadj_bound}
\norm{ \prod_{j=1}^k \left[ \Phi(f_j) + \normord{ \Phi^2 }(h_j) \right] \psi } \leq \frac{10^k \, \Gamma\left( k + \frac{n+1}{2} \right)}{\Gamma\left( \frac{n+1}{2} \right)} \prod_{j=1}^k K(f_j, h_j) \norm{ \psi } \eqend{,}
\end{splitequation}
where $K(f,h)$ is an explicit constant, satisfying $K(\alpha f, \beta h) \leq \max(|\alpha|,|\beta|) K(f,h)$ for all $\alpha, \beta \in \mathbb{R}$.

In particular, for arbitrary $f, h \in \mathcal{S}_\mathbb{R}(\mathbb{R}^d)$, any real-linear combination of $\Phi(f)$ and $\normord{\Phi^2}(h)$ is essentially selfadjoint on $\mathcal{F}_0$.
\end{lemma}
\begin{proof}
The required bound is~\eqref{eq:fixedparticlebound_k}, where the constant $K(f,h)$ is given in Eqs.~\eqref{eq:kalphabeta_def} and~\eqref{eq:ch_def}, and the scaling is given in Eq.~\eqref{eq:k_scaling}.

Essential selfadjointness follows using Nelson's analytic vector theorem~\cite[Thm. X.39]{reedsimon2}: For arbitrary $\alpha,\beta \in \mathbb{R}$, the linear combination $A \defby \alpha \Phi(f) + \beta \normord{\Phi^2}(h) = \Phi(\alpha f) + \normord{\Phi^2}(\beta h)$ is symmetric on $D(A) = \mathcal{F}_0$. Moreover, all vectors $\psi \in \mathcal{F}_0$ are in $D(A^k)$ for all $k$ by the bound~\eqref{eq:wicksadj_bound}. They are also analytic for $A$ since
\begin{splitequation}
\sum_{k = 0}^\infty \frac{s^k}{k!} \norm{ A^k \psi } &\leq \sum_{k = 0}^\infty \frac{s^k}{k!} \frac{\Gamma\left( k + \frac{n+1}{2} \right)}{\Gamma\left( \frac{n+1}{2} \right)} \left[ 10 K(\alpha f, \beta h) \right]^k \norm{ \psi } \\
&= \left[ 1 - 10 s K(\alpha f, \beta h) \right]^{- \frac{n+1}{2}} \norm{ \psi } \eqend{,}
\end{splitequation}
as long as $s < 1/[ 10 K(\alpha f, \beta h) ]$, independently of $\psi$.
\end{proof}
This allows us to define through spectral calculus
\begin{equation}
\label{eq:w1_w2_def}
W_1(f) \defby \exp\left[ \mathi \overline{ \Phi(f) } \right] \eqend{,} \quad W_2(h) \defby \exp\left[ \frac{\mathi}{2} \overline{ \normord{\Phi^2}(h) } \right]
\end{equation}
as unitaries on Fock space $\mathcal{F}$, where the overline indicates the closure of operators which in our case coincides with the unique self-adjoint extension. In particular, $W_1(f) \Omega$ and $W_2(h) \Omega$ are well-defined elements of Fock space, and for sufficiently small $K(f,h)$ the definitions~\eqref{eq:w1_w2_def} coincide with the sum of the corresponding formal power series.

Furthermore, this allows us to prove that the formal relations~\eqref{eq:bch} and~\eqref{eq:bch2} are actually well-defined.
\begin{lemma}
\label{lem:welldef}
For any $f, h \in \mathcal{S}_\mathbb{R}(\mathbb{R}^d)$, the identities
\begin{equation}
\label{eq:commutator1_w2}
W_2(-h) \overline{ \Phi(f) } W_2(h) = \overline{ \Phi\left( T_h f \right) }
\end{equation}
holds on the finite-particle subspace $\mathcal{F}_0$. Moreover, on all of $\mathcal{F}$, it holds that
\begin{equation}
\label{eq:commutator2_w2}
W_2(-h) W_1(f) W_2(h) = W_1(T_h f) \eqend{.}
\end{equation}
Consequently, for $h \in \mathcal{S}_\mathbb{R}(\mathcal{W})$ it holds that $W_2(h) \in \mathcal{A}_\mathcal{W}$.
\end{lemma}
\begin{proof}
Consider a vector $\psi \in \mathcal{F}_0$ containing at most $n$ particles. For the identity \eqref{eq:commutator1_w2}, we first show that $W_2(s h) \psi$ is in the domain of definition of $\overline{ \Phi(f) }$ for suitably small $s > 0$. For this, let
\begin{equation}
W_{2,M}(s h) \defby \sum_{k=M}^\infty \frac{\mathi^k}{2^k k!} \left[ \normord{ \Phi^2 }(s h) \right]^k
\end{equation}
and consider the graph norm
\begin{splitequation}
&\norm{ W_{2,M}(s h) \psi } + \norm{ \Phi(f) W_{2,M}(s h) \psi } \\
&\leq \sum_{k=M}^\infty \frac{s^k}{2^k k!} \left( \norm{ \left[ \normord{ \Phi^2 }(h) \right]^k \psi } + \norm{ \Phi(f) \left[ \normord{ \Phi^2 }(h) \right]^k \psi } \right) \\
&\leq \sum_{k=M}^\infty \frac{(5 s)^k}{k!} \left( \frac{\Gamma\left( k + \frac{n+1}{2} \right)}{\Gamma\left( \frac{n+1}{2} \right)} + \frac{10 \Gamma\left( k + \frac{n+3}{2} \right)}{\Gamma\left( \frac{n+1}{2} \right)} K(f, 0) \right) \left[ K(0, h) \right]^k \norm{ \psi } \\
&\leq \left[ 5 s K(0, h) \right]^M \frac{1 + 10 K(f, 0)}{\Gamma\left( \frac{n+1}{2} \right)} \sum_{k=0}^\infty a_k \left[ 5 s K(0, h) \right]^k \norm{ \psi }
\end{splitequation}
with
\begin{equation}
a_k = \frac{\Gamma\left( k + M + \frac{n+3}{2} \right)}{(k+M)!} \eqend{,}
\end{equation}
where in the second inequality we used the bounds~\eqref{eq:wicksadj_bound} and in the last equality shifted $k \to k+M$. Since for the ratio of coefficients we have
\begin{equation}
\frac{a_{k+1}}{a_k} = \frac{k + M + \frac{n+3}{2}}{k+M+1} \to 1 \quad (k+M \to \infty) \eqend{,}
\end{equation}
the sum is absolutely convergent (independently of $M$) if $5 s K(0,h) < 1$. Assuming this condition, we also obtain
\begin{equation}
\lim_{M \to \infty} \Bigl( \norm{ W_{2,M}(s h) \psi } + \norm{ \Phi(f) W_{2,M}(s h) \psi } \Bigr) = 0 \eqend{,}
\end{equation}
and hence the partial sums converge in the graph norm to zero. Since $\overline{ \Phi(f) }$ is closed, it follows that $W_2(s h) \psi$ is in its domain of definition (at least) if $5 s K(0,h) < 1$.

We can then check that
\begin{splitequation}
\norm{ W_{2,M}(-s h) \Phi(f) W_{2,N}(s h) \psi } &\leq \left[ 5 s K(0,h) \right]^{M+N} \frac{10 K(f,0)}{\Gamma\left( \frac{n+1}{2} \right)} \norm{ \psi } \\
&\quad\times \sum_{k,\ell=0}^\infty \frac{\Gamma\left( k+\ell+M+N+ \frac{n+3}{2} \right)}{(k+M)! (\ell+N)!} \left[ 5 s K(0,h) \right]^{k+\ell}
\end{splitequation}
by a computation completely analogous to the one for the graph norm. Since all terms in the double sum are positive, by \Cref{lem:pringsheim} in \Cref{app:pringsheim} the double sum converges if any rearrangement of it converges, and gives the same limiting value in all cases. In particular we may take the diagonal rearrangement $R(r) = \{ (k,\ell) \colon k + \ell = r \}$, and compute
\begin{splitequation}
&\sum_{(k,\ell) \in R(r)} \frac{\Gamma\left( k+\ell+M+N+ \frac{n+3}{2} \right)}{(k+M)! (\ell+N)!} \left[ 5 s K(0,h) \right]^{k+\ell} \\
&\quad= \left[ 5 s K(0,h) \right]^r \sum_{k=0}^r \frac{\Gamma\left( r+M+N+ \frac{n+3}{2} \right)}{(k+M)! (r-k+N)!} \\
&\quad= \left[ 5 s K(0,h) \right]^r \frac{\Gamma\left( r+M+N+ \frac{n+3}{2} \right)}{\Gamma(r+M+N+1)} \sum_{k=0}^r \binom{M+N+r}{M+k} \\
&\quad\leq \left[ 5 s K(0,h) \right]^r \frac{\Gamma\left( r+M+N+ \frac{n+3}{2} \right)}{\Gamma(r+M+N+1)} \sum_{k=0}^{M+N+r} \binom{M+N+r}{k} \\
&\quad= \left[ 5 s K(0,h) \right]^r \frac{\Gamma\left( r+M+N+ \frac{n+3}{2} \right)}{\Gamma(r+M+N+1)} 2^{M+N+r} \eqend{,}
\end{splitequation}
such that
\begin{splitequation}
\norm{ W_{2,M}(-s h) \Phi(f) W_{2,N}(s h) \psi } &\leq \left[ 5 s K(0,h) \right]^{M+N} \frac{10 K(f,0)}{\Gamma\left( \frac{n+1}{2} \right)} \norm{ \psi } \\
&\quad\times \sum_{r=0}^\infty \sum_{(k,\ell) \in R(r)} \frac{\Gamma\left( k+\ell+M+N+ \frac{n+3}{2} \right)}{(k+M)! (\ell+N)!} \left[ 5 s K(0,h) \right]^{k+\ell} \\
&\leq \left[ 10 s K(0,h) \right]^{M+N} \frac{10 K(f,0)}{\Gamma\left( \frac{n+1}{2} \right)} \norm{ \psi } \\
&\quad\times \sum_{r=0}^\infty \left[ 10 s K(0,h) \right]^r \frac{\Gamma\left( r+M+N+ \frac{n+3}{2} \right)}{\Gamma(r+M+N+1)} \eqend{.}
\end{splitequation}
The sum over $r$ converges by the ratio test (independently of $M$ and $N$) if $10 s K(0,h) < 1$, and under the same condition, the bound vanishes as $M,N \to \infty$ independently of each other. Therefore, we have
\begin{equation}
\lim_{M,N \to \infty} \left( \sum_{k=0}^M \frac{(-\mathi)^k}{2^k k!} \left[ \normord{ \Phi^2 }(s h) \right]^k \right) \Phi(f) \left( \sum_{\ell=0}^N \frac{\mathi^\ell}{2^\ell \ell!} \left[ \normord{ \Phi^2 }(s h) \right]^\ell \right) \psi = W_2(-s h) \Phi(f) W_2(s h) \psi
\end{equation}
in norm, and the right-hand side is an analytic function of $s$ (at least) for $s < 1/[10 K(0,h)]$.

Because $T_h = \exp(t_h)$~\eqref{eq:th2} is a bounded operator on the one-particle Hilbert space by~\Cref{lem:thdef} and $t_{s h} = s t_h$ by definition~\eqref{eq:th1}, it follows that
\begin{equation}
K(T_{sh}f,0) = \norm{T_{sh} f} = \norm{\exp(s t_h) f} \leq \exp\left( \abs{s} \norm{t_h} \right) \norm{f} \leq c_h^\abs{s} \norm{f}
\end{equation}
for some $c_h < \infty$. Using again the estimate \eqref{eq:wicksadj_bound}, we thus see that $\Phi(T_{s h} f) \psi$ is an analytic function of $s$ in a neighborhood of the origin. It follows that the difference
\begin{equation}
W_2(-s h) \Phi(f) W_2(s h) \psi - \Phi(T_{s h} f) \psi
\end{equation}
is an analytic function of $s$ in a neighborhood of the origin, and since each of its Taylor coefficients vanishes by construction, the function vanishes in a neighborhood of the origin. By the uniqueness of analytic continuation, it vanishes everywhere it is defined, giving identity \eqref{eq:commutator1_w2}, holding on $\mathcal{F}_0$.

For identity \eqref{eq:commutator2_w2}, the proof is analogous: by construction, all Taylor coefficients of
\begin{equation}
W_2(-s h) W_1(t f) W_2(s h) \psi - W_1(t \, T_{s h} f) \psi
\end{equation}
at the origin $s = t = 0$ vanish. Using that $T_h = \exp(t_h)$ is a bounded operator on the one-particle Hilbert space by~\Cref{lem:thdef} and $t_{s h} = s t_h$, and employing \Cref{lem:wicksadj} and the diagonal rearrangement in \Cref{lem:pringsheim}, one shows that both sides are analytic functions of $s$ and $t$ in a neighborhood of the origin, and by the uniqueness of analytic continuation they agree everywhere. Since \eqref{eq:commutator2_w2} involves only unitary, thus bounded, operators, the identity extends from $\mathcal{F}_0$ to all of $\mathcal{F}$.

Finally, consider $h \in \mathcal{S}_\mathbb{R}(\mathcal{W})$ and $f \in \mathcal{S}_\mathbb{R}(\mathcal{W}^\top)$ with the opposite wedge region $\mathcal{W}^\top$, consisting of all points spacelike related to the closure of $\mathcal{W}$. By Haag duality, the Weyl algebra of the opposite wedge is the commutant of the wedge algebra: $\mathcal{A}(\mathcal{W}^\top) = \mathcal{A}'(\mathcal{W})$, see e.g. \cite[Thm.~2.14]{Hislop1986} or \cite[Prop.~6]{Verch1993}. Furthermore, since the commutator function is causal we obtain $h (\Delta \ast f) = 0$ and thus $T_h f = f$. The identity~\eqref{eq:commutator2_w2} then shows that
\begin{equation}
[ W_1(f), W_2(h) ] = W_2(h) W_1(T_h f) - W_2(h) W_1(f) = 0 \eqend{,}
\end{equation}
such that $W_2(h)$ commutes with all generators of $\mathcal{A}'(\mathcal{W})$. By von Neumann's bicommutant theorem, it is thus an element of $\mathcal{A}''(\mathcal{W}) = \mathcal{A}(\mathcal{W})$.
\end{proof}

As a straightforward corollary, we obtain
\begin{theorem}
\label{thm:uimpl}
The Bogoliubov transformation $\alpha_{T_h}$ given by~\eqref{eq:bogoliubov_w1} is unitarily implement\-able, namely we have
\begin{equation}
\label{eq:exprel}
\alpha_{T_h}( W_1(f) ) = W_2(h)^\dagger W_1(f) W_2(h) = W_1(T_h f) \eqend{.}
\end{equation}
As a consequence, $\omega$ and $\omega_{T_h} = \omega \circ \alpha_{T_h}$ are in the same folium. That is, they can be represented as vector states in the same Fock representation.
\end{theorem}
\begin{proof}
The identity \eqref{eq:exprel} follows directly from the definition~\eqref{eq:bogoliubov_w1} of $\alpha_{T_h}$ and \Cref{lem:welldef}.

In the Fock representation, $\omega$ is represented by the vector state $\ket{\Omega}$, and from the relation~\eqref{eq:exprel} it follows that $\omega_{T_h}$ is represented by $W_2(h) \ket{\Omega}$, such that $\omega_{T_h} = \omega \circ \operatorname{Ad}_{W_2(h)}$ with $\operatorname{Ad}_{U}(A)\defby U^\dagger A U$ for a unitary $U$.
\end{proof}

We conclude that unitary excitations associated with the smeared Wick square acting on the Minkowski vacuum give a class of well-defined quasi-free states, which can be produced by switching on and off a quadratic Hamiltonian (more or less by definition), and can be represented in the vacuum Fock representation (as we have proven). These states can be taken as a straightforward generalization of the coherent states obtained by acting with unitary excitations of the smeared field itself and, as was motivated by the previous sections, can be understood as \emph{locally squeezed states} in spacetime.

\subsection{Local squeezing is Hadamard-preserving and produces finite energy and finite wedge entropy}
\label{sec:finiteentropy}

As our main result, we will show that local squeezing has desirable properties of a physically realizable operation. In particular, we will show that local squeezing of the Minkowski vacuum state produces a smooth change of its correlations and its energy density. That is, squeezing preserves the Hadamard condition on the state, which is a requirement for physically reasonable states in QFT~\cite{FV13}. It follows that also the change of total energy and the relative entropy of the squeezed state with respect to the Minkowski vacuum are finite, the latter under the condition that the squeezing is restricted to a wedge.

To begin with, we will need suitable estimates. For this, recall the Schwartz seminorms
\begin{equation}
\label{eq:schwartz_seminorm}
\norm{ f }_{K,\alpha,\beta} \defby \sup_{x \in K} \abs{ x^\alpha \partial^\beta f(x) } \eqend{,} \quad \norm{ f }_{K,m,n} \defby \sup_{\abs{\alpha} \leq m, \abs{\beta} \leq n} \norm{ f }_{K,\alpha,\beta} \eqend{,}
\end{equation}
where $\alpha, \beta$ are multiindices and $m,n \in \mathbb{N}_0$. For distributions $A, B \colon \mathcal{S}(\mathbb{R}^d \times \mathbb{R}^d) \to \mathbb{C}$ with integral kernels $A_2$, $B_2$, we introduce the transpose and composition by
\begin{equation}
A_2^\top(x,y) \defby A_2(y,x) \eqend{,} \quad (A \circ B)_2(x,y) = \int A_2(x,z) B_2(z,y) \total^d z \eqend{,}
\end{equation}
where $A \circ B$ is only defined if the distributions $A$ and $B$ satisfy a suitable wave front set and decay condition~\cite[Thm. 8.2.14]{Hormander1}. Moreover, we employ $h_2$ as notation for the multiplication kernel
\begin{equation}
h_2(x,y) \defby h(x) \delta(x-y) \eqend{.}
\end{equation}
The commutator function $\Delta$~\eqref{eq:delta_def}, the symmetric bilinear form $\mu$~\eqref{eq:mu_def} and the operator $t_h$~\eqref{eq:th1} can then be written succintly as
\begin{equation}
\label{eq:recall}
\mathi \Delta = \omega_2 - \omega_2^\top \eqend{,} \quad \mu = \frac{1}{2} \left( \omega_2 + \omega_2^\top \right) \eqend{,} \quad t_h = h_2 \circ \Delta \eqend{.}
\end{equation}

We first show
\begin{lemma}
\label{lem:vacestimate}
For $n \in \mathbb{N}$, consider $\sigma \defby (\sigma_1,\ldots,\sigma_{n+1}) \in \{+1,-1\}^{n+1}$ and define
\begin{equation}
\label{eq:defword}
W_{\sigma,h}^{(n)} \defby w_{\sigma_1} \circ h_2 \circ w_{\sigma_2} \circ h_2 \circ \cdots \circ h_2 \circ w_{\sigma_{n+1}} \eqend{,}
\end{equation}
where we set $w_+ \defby \omega_2$ and $w_- \defby \omega_2^\top$. Assume that $\sigma$ is mixed, i.e., $\sigma \neq (+1,\ldots,+1)$ and $\sigma \neq (-1,\ldots,-1)$. Then $W_{\sigma,h}^{(n)}$ has a smooth kernel, and for all
$j,k \in \mathbb{N}_0$ and every temporally compact set $K \subset \mathbb{R}^d \times \mathbb{R}^d$, there exist constants $C_{K,j,k} \geq 0$ and
$\ell,m \in \mathbb{N}_0$ which are independent of $n$, $\sigma$ and $h$, such that
\begin{equation}
\norm{ W_{\sigma,h}^{(n)} }_{K,j,k} \leq \left( C_{K,j,k} \norm{ h }_{\mathbb{R}^d,\ell,m} \right)^n \eqend{.}
\end{equation}
\end{lemma}

\begin{proof}
Eq.~\eqref{eq:2ptfunc} shows that the integral kernel of $w_\sigma$ reads
\begin{equation}
w_\sigma(x,y) = \int \mathe^{\mathi \sigma p(x-y)} \Bigr\rvert_{p^0 = \omega_\vec{p}} \total \mu(\vec{p}) \eqend{,}
\end{equation}
such that we obtain
\begin{splitequation}
W_{\sigma,h}^{(n)}(x,y) &= \int\dotsi\int \mathe^{\mathi \sigma_1 p_1 x - \mathi \sigma_{n+1} p_{n+1} y} \\
&\qquad\times \prod_{j=1}^n \tilde{h}(\sigma_j p_j - \sigma_{j+1} p_{j+1}) \Bigr\rvert_{p_j^0 = \omega_{\vec{p}_j}} \prod_{j=1}^{n+1} \total \mu(\vec{p}_j) \eqend{.}
\end{splitequation}
We now need to bound Schwartz norms of $W_{\sigma,h}^{(n)}(x,y)$. Derivatives with respect to $x$ or $y$ simply result in polynomial factors in
$p_1$ and $p_{n+1}$. With $x = (t,\vec{x})$ and $y = (s,\vec{y})$, since $K$ is temporally compact, multiplication by powers of $t$ or $s$ can be estimated by a constant depending on $K$. Multiplication by a power of $\vec{x}$ can be expressed via
\begin{equation}
\vec{x}^j \mathe^{\mathi \sigma_1 p_1 x} = \frac{1}{\mathi \sigma_1} \frac{\partial}{\partial \vec{p}_1^j} \mathe^{\mathi \sigma_1 p_1 x} + \frac{\vec{p}_1^j}{\omega_{\vec{p}_1}} t \, \mathe^{\mathi \sigma_1 p_1 x} \eqend{,}
\end{equation}
and the partial derivative with respect to $\vec{p}_1$ can then be integrated by parts, acting either on the Fourier transforms $\tilde{h}$ or on the measure factor $1/(2 \omega_{\vec{p}_1})$, where it results in a factor $- \vec{p}_1^j/(2 \omega_{\vec{p}_1}^3)$. After taking the absolute value, we further estimate $\abs{ \vec{p}^j/\omega_\vec{p} } \leq 1$ and $1/\omega_\vec{p} \leq 1/m$. Analogous results holds for multiplication by a power of $\vec{y}$ and involve $\vec{p}_{n+1}$.

Taking all of this into account, we obtain
\begin{equation}
\norm{ W_{\sigma,h}^{(n)} }_{K,\alpha,\beta} \leq C \norm{ a_\sigma^{(n+1)} }_{\abs{\beta},\abs{\alpha}} \eqend{,}
\end{equation}
where $C > 0$ is a constant depending on $\alpha$, $\beta$ and $K$, but not on $n$, $\sigma$ or $h$,
\begin{equation}
a_\sigma^{(n+1)}(\vec{p}_1,\vec{p}_{n+1}) \defby \int\cdots\int \prod_{j=1}^n \tilde{h}(\sigma_j p_j - \sigma_{j+1} p_{j+1}) \Bigr\rvert_{p_j^0 = \omega_{\vec{p}_j}} \prod_{j=2}^n \total \mu(\vec{p}_j)
\end{equation}
and the norm $\norm{ \cdot }_{k,\ell}$ is a Schwartz-type norm defined in Eq.~\eqref{eq:bound_kernel_norm} with $k$ measuring powers of momenta and $\ell$ orders of derivatives with respect to them. By \Cref{lemma:bound_kernel}, there exist constants $C' > 0$ and $\ell', m' \in \mathbb{N}_0$ depending only on $\abs{\alpha}, \abs{\beta}$ such that
\begin{equation}
\norm{ a_\sigma^{(n+1)} }_{\abs{\beta},\abs{\alpha}} \leq \left[ C' \norm{ h }_{\mathbb{R}^d,\ell',m'} \right]^n \eqend{.}
\end{equation}
Hence, arbitrary Schwartz norms of the integral kernel $W_{\sigma,h}^{(n)}(x,y)$ are finite, and the conclusion follows by taking suprema over all values of the multiindices:
\begin{equation}
\{ \ell,m,C_{K,j,k} \} = \sup_{\abs{\alpha} \leq j, \abs{\beta} \leq k} \{ \ell',m', \max(1,C) C' \} \eqend{.}
\end{equation}
\end{proof}

With this, we can prove our main result.
\begin{theorem}
\label{thm:finite}
Let $\omega$ be the vacuum state for the massive scalar field in Minkowski spacetime with $d = D+1$ dimensions and let $\omega_2$ denote its two-point function.
Then, for any $h \in \mathcal{S}_\mathbb{R}(\mathbb{R}^d)$, the distributions $\omega_{T_h,2} - \omega_2$ and $\omega_{T_h}(\normord{T^{00}})$ have smooth function kernels. In particular, $\omega_{T_h}$ is a quasi-free Hadamard state.

Moreover, $\omega_{T_h}$ has finite energy relative to $\omega$:
\begin{equation}
\int \abs{ \omega_{T_h}\left( \normord{T^{00}}(0,\vec{x}) \right) } \total^D \vec{x} < \infty \eqend{.}
\end{equation}
If in addition $\supp h \subset \mathcal{W}$ for the wedge $\mathcal{W}$, then the relative entropy between $\omega_{T_h}$ and $\omega$ in the wedge algebra $\mathcal{A}_\mathcal{W}$ is also finite:
\begin{equation}
S_\mathrm{rel}( W_2(h) \Omega \Vert \Omega; \mathcal{A}_\mathcal{W}) < \infty \eqend{.}
\end{equation}
\end{theorem}
\begin{proof}
By \Cref{lem:sympl}, $T_h$ is a (bounded) linear symplectic transformation of $(\mathcal{S}_{\mathbb{R}}/\ker \Delta, \Delta)$. It follows that
\begin{equation}
\omega_{T_h,2}(f,g) - \omega_2(f,g) = \omega_2(T_h f, T_h g) - \omega_2(f,g) = \mu(T_h f, T_h g) - \mu(f,g) \bydef R_h(f,g) \eqend{,}
\end{equation}
where the commutator function $\Delta$ canceled and only the symmetric bilinear form $\mu$ remained. Expanding $R_h$ in powers of $h$ by expanding $T_h = \exp(t_h)$ and inserting Eq.~\eqref{eq:recall}, we obtain
\begin{equation}
R_h = \sum_{n=1}^\infty R_h^{(n)}
\end{equation}
with
\begin{equation}
R_h^{(n)} \defby \frac{1}{2} \sum_{m=0}^n \frac{\mathi^n (-1)^m}{(n-m)! m!} \left[ ( \omega_2 - \omega_2^\top ) \circ h_2 \right]^{\circ (n-m)} \circ ( \omega_2 + \omega_2^\top ) \circ \left[ h_2 \circ ( \omega_2 - \omega_2^\top ) \right]^{\circ m} \eqend{.}
\end{equation}
Expanding each factor $\omega_2 - \omega_2^\top$ and the middle factor
$\omega_2 + \omega_2^\top$, we see that every summand has the form $W_{h,\sigma}^{(n)}$ for some mixed $\sigma \in \{+1,-1\}^{n+1}$ as defined in \eqref{eq:defword}. Indeed, the coefficient of $W_{h,(+1,\ldots,+1)}^{(n)}$ is given by
\begin{equation}
\frac{1}{2} \sum_{m=0}^n \frac{\mathi^n (-1)^m}{(n-m)! m!} = \frac{\mathi^n}{2 n!} \sum_{m=0}^n (-1)^m \binom{n}{m} = 0 \eqend{,}
\end{equation}
and the coefficient of $W_{h,(-1,\ldots,-1)}^{(n)}$ vanishes by the same computation.

We can thus use \Cref{lem:vacestimate} with $j = k = 0$ to see that every summand has a smooth kernel. Moreover, since the bounds on $W_{h,\sigma}^{(n)}$ are independent of $\sigma$, for every temporally compact set $K \subset \mathbb{R}^d \times \mathbb{R}^d$ we obtain the bound
\begin{splitequation}
\norm{ R_h^{(n)} }_{K,j,k} &\leq 2^n \sum_{m=0}^n \frac{1}{(n-m)! m!} \max_{\text{mixed }\sigma} \norm{ W_{\sigma,h}^{(n)} }_{K,j,k} = \frac{4^n}{n!} \max_{\text{mixed }\sigma} \norm{ W_{\sigma,h}^{(n)} }_{K,j,k} \\
&\leq \frac{4^n}{n!} \left( C_{K,j,k} \norm{ h }_{\mathbb{R}^d,\ell,m} \right)^n \eqend{,}
\end{splitequation}
where the constants $C_{K,j,k}$, $\ell$ and $m$ do not depend on $n$. It follows that the bound
\begin{equation}
\label{eq:Rh_bound}
\norm{ R_h }_{K,j,k} \leq \sum_{n=1}^\infty \norm{ R_h^{(n)} }_{K,j,k} \leq \exp\left[ 4 C_{K,j,k} \norm{ h }_{\mathbb{R}^d,\ell,m} \right]
\end{equation}
holds for each Schwartz norm, such that $R_h$ is smooth. Since $\omega_{T_h,2} = \omega_2 + R_h$, it follows that the wavefront set of $\omega_{T_h,2}$ is equal to the wavefront set of $\omega_2$. Moreover, we showed in \Cref{thm:bogtrafo} that $\omega_{T_h}$ is quasifree, and thus $\omega_{T_h}$ is a quasifree Hadamard state.

We now consider the energy density and total energy. Let $D^{00}$ denote the standard point-splitting differential operator for the energy density:
\begin{equation}
D^{00} f(x,y) \defby \frac{1}{2} \sum_{\nu=0}^D \partial_x^\nu \partial_y^\nu f(x,y) + \frac{1}{2} m^2 f(x,y) \eqend{.}
\end{equation}
The expectation value of the normal-ordered stress tensor can be defined via point-splitting \cite{Moretti2001,HollandsWald2004}, and we obtain
\begin{equation}
\label{eq:eeexp}
\omega_{T_h}\left( \normord{T^{00}}(x) \right) = \left[ D^{00} \, \omega_{T_h}\left( \normord{ \Phi(x) \Phi(y) } \right) \right]_{y=x} = \left[ D^{00} R_h(x,y) \right]_{y=x} \eqend{.}
\end{equation}
Since $R_h$ is smooth, $D^{00} R_h(x,y)$ is also smooth, and the restriction to the diagonal of a smooth function of two variables is again smooth. Therefore, the energy density $\omega_{T_h}\left( \normord{T^{00}} \right)$ is also smooth.

For the total energy, we use the bound~\eqref{eq:Rh_bound} with $j = d$, $k=2$ for $K = \left( \{0\} \times \mathbb{R}^{D} \right)^{\times 2}$ together with the explicit form of the Schwartz seminorms~\eqref{eq:schwartz_seminorm}. This gives
\begin{equation}
\sup_{\abs{\alpha} \leq d, \abs{\beta} \leq 2} \sup_{(\vec{x},\vec{y}) \in \mathbb{R}^D \times \mathbb{R}^D} \abs{ (x,y)^\alpha \partial^\beta R_h(x,y) } \leq C_h
\end{equation}
with a constant $C_h$, and hence
\begin{equation}
\abs{ \omega_{T_h}\left( \normord{T^{00}}(0,\vec{x}) \right) } = \abs{ \left[ D^{00} R_h(x,y) \right]_{\vec{y} = \vec{x}, y^0 = x^0 = 0} } \leq \frac{C_h}{\max( 1, \abs{\vec{x}} )^d} \eqend{.}
\end{equation}
Since the right-hand side of this bound is integrable over the time-zero Cauchy hypersurface $\mathbb{R}^D$, the total energy is finite.

Analogously, choosing $j = d+1$ we obtain
\begin{equation}
\abs{ x^1 \, \omega_{T_h}\left( \normord{T^{00}}(0,\vec{x}) \right) } \leq \frac{C_h}{\max( 1, \abs{\vec{x}} )^d} \eqend{,}
\end{equation}
and the integral $\displaystyle\int_{\mathbb{R}^D} x^1 \, \omega_{T_h}\left( \normord{T^{00}}(0,\vec{x}) \right) \total^D \vec{x}$ is finite. Since for $\supp h \subset \mathcal{W}$ with the canonical right wedge $\mathcal{W}$, the unitary $W_2(h)$ implementing $T_h$ by \Cref{thm:uimpl} is an element of the wedge algebra $\mathcal{A}_\mathcal{W}$ by \Cref{lem:welldef}, the relative entropy between $\omega_{T_h}$ and $\omega$ is given by the formula~\eqref{eq:relentropy}, which is the restriction of this integral to $x^1 > 0$, multiplied by $2 \pi$. Hence, the relative entropy is also finite, and using Poincaré covariance we see that the result holds also for an arbitrary wedge.
\end{proof}

\section{Conclusions}
\label{sec:conclusion}

We have proposed two classes of states for quantum field theory which have so far received little attention in the literature, and which we call \emph{locally squeezed states}. These arise by acting with a unitary operator, formed from the exponential of a local generator quadratic in the basic field (e.g., a Hamiltonian) on the Fock vacuum state of the real massive scalar field in Minkowski spacetime. Here a generator is viewed as \emph{local} if it arises as the quantization of a point-local functional. Depending on the formalism, canonical or covariant, we distinguish local squeezing in space or spacetime. Despite the fact that both notions give rise to well-defined quasi-free states, the relative entropy between a locally squeezed state in space and the vacuum generally diverges, however small the squeezing is. However, as we have shown local squeezing in spacetime yields both a Hadamard state and a finite relative entropy.

Since most alternative measures of state difference like logarithmic dominance, mutual information and the modular entanglement measure dominate the relative entropy \cite{HS18}, also these will be infinite for space-local squeezing. The same holds for generalizations of relative entropy, such as the (Petz--)R{\'e}nyi relative entropy or sandwiched and geometric quantum R{\'e}nyi divergences if $\alpha > 1$, see Refs.~\cite{petz1985,wildewinteryang2014,muellerlennertetal2013,beigi2013,hiai2018,matsumoto2018,jencova2021} and references therein. On the other hand, if $\alpha \in [0,1)$, the relative entropy is an upper bound for the Petz--R{\'e}nyi relative entropy (see for example~\cite{froebsangaletti2025}), and it would be interesting to see if this entropy is finite for such $\alpha$ for space-local squeezing.

While our construction focusses on the simplest setting (the real massive scalar field on Minkowski spacetime in the usual Fock vacuum representation), it is clear that the construction can be generalized to other types of fields in other types of representations. For example, it would be interesting to see how squeezing works for fermions, where relative entropy and R{\'e}nyi relative entropy for Gaussian states have been computed recently~\cite{galandamuchverch2023,FinsterLottnerMuchMurro:2024,FinsterCausalFermions:2025,FinsterMuch:2025}. Moreover, assuming suitable results on the selfadjointness of the Wick square, such as the ones derived in \cite{San12}, it could be further generalized to curved spacetimes. In particular, we expect that local squeezing can be defined on globally hyperbolic spacetimes for arbitrary quasi-free states.

One interesting application of relative entropy for squeezed states would be the computation of the entropy produced by the expansion of our universe, which has also been studied numerically (including a UV cutoff)~\cite{boutivaspastrastetradis2023,boutivaskatsinispastrastetradis2024,boutivaskatsinispastrastetradis2025a,boutivaskatsinispastrastetradis2025b}. In particular, the Bunch--Davies vacuum state can be understood as a squeezed state with respect to the Minkowski vacuum state, such that the relative entropy between both gives the contribution to the entropy of the Bunch--Davies vacuum that is due to the expansion. This squeezing is non-local, and will need an extension of our results.

\acknowledgments

We thank Henning Bostelmann, Ko Sanders, and Leonardo Sangaletti for valuable discussions and remarks. We also thank an anonymous referee for pointing out a crucial error in an earlier computation of the relative entropy and for a very thorough review of our manuscript.

D.~Cadamuro and M.~B.~Fr{\"o}b have been partially funded by the Deutsche Forschungsgemeinschaft (DFG, German Research Foundation) --- project no. 396692871 within the Emmy Noether grant CA1850/1-1.

M.~B.~Fr{\"o}b thanks Miguel Navascu{\'e}s and the IQOQI, as well as N. Tetradis and D. Katsinis for their generous hospitality during visits to Vienna and Athens.

J.~Mandrysch acknowledges funding by the Austrian Science Fund (FWF) [10.55776/\linebreak COE1] and the European Union NextGenerationEU within the quantA Core project ``Local Operations in Quantum fields''. He thanks the Institute for Theoretical Physics at the University of Leipzig for hospitality during which a part of this work was conducted. The visit was funded by the COST Action ``Relativistic Quantum Information'' CA23115.

\section*{Conflict of interest statement}

On behalf of all authors, the corresponding author states that there is no conflict of interest.

\appendix

\section{Computations for squeezing in space}\label{app:sqspace}

In this section we determine the squeezed fields defined in \eqref{eq:locsqfields} and normal-ordered products thereof. Using the canonical commutation relations \eqref{eq:ccr} and $[\normord{A^2}(\vec{x}),B] = \lim_{\vec{y} \to \vec{x}} [A(\vec{x}) A(\vec{y}),B]$ we have
\begin{equations}[eq:comm_a]
[ A(k), \phi(\vec{x}) ] &= - \mathi k_0(\vec{x}) \phi(\vec{x}) - \mathi k_\pi(\vec{x}) \pi(\vec{x}) \eqend{,} \\
[ A(k), \pi(\vec{x}) ] &= \mathi k_\phi(\vec{x}) \phi(\vec{x}) + \mathi k_0(\vec{x}) \pi(\vec{x}) \eqend{,}
\end{equations}
and from this
\begin{equations}[eq:comm_ak_0]
[ A(k), \phi(\vec{x}) ]_n &= f_n(\vec{x}) \phi(\vec{x}) + g_n(\vec{x}) \pi(\vec{x}) \eqend{,} \\
[ A(k), \pi(\vec{x}) ]_n &= h_n(\vec{x}) \phi(\vec{x}) + l_n(\vec{x}) \pi(\vec{x})
\end{equations}
with certain coefficient functions $f_n,g_n,h_n,l_n$ for the iterated commutators
\begin{equation}
\label{eq:itcomm}
[ A, B ]_n = [ A, [ A, B ]_{n-1} ] \eqend{,} \quad [A,B]_0 = B \eqend{.}
\end{equation}
From the recursive definition~\eqref{eq:itcomm} we obtain the recursion relations
\begin{equations}
\begin{pmatrix} f_n(\vec{x}) \\ g_n(\vec{x}) \end{pmatrix} &= \mathi \begin{pmatrix} - k_0(\vec{x}) & k_\phi(\vec{x}) \\ - k_\pi(\vec{x}) & k_0(\vec{x}) \end{pmatrix} \begin{pmatrix} f_{n-1}(\vec{x}) \\ g_{n-1}(\vec{x}) \end{pmatrix} \eqend{,} \label{eq:matrixdef1}\\
\begin{pmatrix} h_n(\vec{x}) \\ l_n(\vec{x}) \end{pmatrix} &= \mathi \begin{pmatrix} - k_0(\vec{x}) & k_\phi(\vec{x}) \\ - k_\pi(\vec{x}) & k_0(\vec{x}) \end{pmatrix} \begin{pmatrix} h_{n-1}(\vec{x}) \\ l_{n-1}(\vec{x}) \end{pmatrix} \label{eq:matrixdef2}
\end{equations}
and the initial conditions
\begin{equation}
f_0(\vec{x}) = l_0(\vec{x}) = 1 \eqend{,} \quad g_0(\vec{x}) = h_0(\vec{x}) = 0 \eqend{.}
\end{equation}

We note that the matrix appearing in Eqs.~\eqref{eq:matrixdef1} and \eqref{eq:matrixdef2} squares to
\begin{equation}
    \epsilon(k,\vec{x}) \defby k_0(\vec{x})^2 - k_\phi(\vec{x}) k_\pi(\vec{x})
\end{equation}
times the identity. Thus, it is easy to solve these relations and we obtain
\begin{equations}
\begin{pmatrix} f_{2n}(\vec{x}) \\ g_{2n}(\vec{x}) \end{pmatrix} &= (-1)^n \epsilon(k,\vec{x})^n \begin{pmatrix} 1 \\ 0 \end{pmatrix}\eqend{,}
& \quad
\begin{pmatrix} f_{2n+1}(\vec{x}) \\ g_{2n+1}(\vec{x}) \end{pmatrix} & = (-1)^n\epsilon(k,\vec{x})^n \mathi \begin{pmatrix} - k_0(\vec{x}) \\   -k_\pi(\vec{x}) \end{pmatrix}  \eqend{,} \\
\begin{pmatrix} h_{2n}(x) \\ l_{2n}(x) \end{pmatrix} &= (-1)^n \epsilon(k,\vec{x})^n \begin{pmatrix} 0 \\ 1 \end{pmatrix} \eqend{,}
& \quad
\begin{pmatrix} h_{2n+1}(x) \\ l_{2n+1}(x) \end{pmatrix} &= (-1)^n \epsilon(k,\vec{x})^n \mathi \begin{pmatrix} k_\phi(\vec{x}) \\ k_0(\vec{x}) \end{pmatrix} \eqend{.}
\end{equations}

Putting everything together, we find
\begin{equations}
\begin{split}
\phi_S(\vec{x}) &= \sum_{n=0}^\infty \frac{(-\mathi)^n}{n!} \left[ f_n(\vec{x}) \phi(\vec{x}) + g_n(\vec{x}) \pi(\vec{x}) \right]\\
&= k_{c,-}(\vec{x}) \phi(\vec{x}) - k_{s,\pi}(\vec{x}) \pi(\vec{x}) \eqend{,}
\end{split}\\
\begin{split}
\pi_S(\vec{x}) &= \sum_{n=0}^\infty \frac{(-\mathi)^n}{n!} \left[ h_n(\vec{x}) \phi(\vec{x}) + l_n(\vec{x}) \pi(\vec{x}) \right]\\
&= k_{c,+}(\vec{x}) \pi(\vec{x}) + k_{s,\phi}(\vec{x}) \phi(\vec{x}) \eqend{.}
\end{split}
\end{equations}

For normal-ordered products, we obtain from this
\begin{equations}
\begin{split}
\label{eq:nordprod1}
S(k)^\dagger \normord{ \phi(\vec{x}) \phi(\vec{y}) } S(k) &= \phi_S(\vec{x}) \phi_S(\vec{y}) - \expect{ \phi(\vec{x}) \phi(\vec{y}) } \\
&= k_{c,-}(\vec{x}) k_{c,-}(\vec{y}) \normord{ \phi(\vec{x}) \phi(\vec{y}) } + k_{s,\pi}(\vec{x}) k_{s,\pi}(\vec{y}) \normord{ \pi(\vec{x}) \pi(\vec{y}) } \\
&\quad- k_{c,-}(\vec{x}) k_{s,\pi}(\vec{y}) \normord{ \phi(\vec{x}) \pi(\vec{y}) } - k_{s,\pi}(\vec{x}) k_{c,-}(\vec{y}) \normord{ \pi(\vec{x}) \phi(\vec{y}) } \\
&\quad+ \left[ k_{c,-}(\vec{x}) k_{c,-}(\vec{y}) - 1 \right] \expect{ \phi(\vec{x}) \phi(\vec{y}) } + k_{s,\pi}(\vec{x}) k_{s,\pi}(\vec{y}) \expect{ \pi(\vec{x}) \pi(\vec{y}) } \\
&\quad- k_{c,-}(\vec{x}) k_{s,\pi}(\vec{y}) \expect{ \phi(\vec{x}) \pi(\vec{y}) } - k_{s,\pi}(\vec{x}) k_{c,-}(\vec{y}) \expect{ \pi(\vec{x}) \phi(\vec{y}) } \eqend{,}
\end{split} \\
\begin{split}
\label{eq:nordprod2}
S(k)^\dagger \normord{ \pi(\vec{x}) \pi(\vec{y}) } S(k) &= \pi_S(\vec{x}) \pi_S(\vec{y}) - \expect{ \pi(\vec{x}) \pi(\vec{y}) } \\
&= k_{c,+}(\vec{x}) k_{c,+}(\vec{y}) \normord{ \pi(\vec{x}) \pi(\vec{y}) } + k_{s,\phi}(\vec{x}) k_{s,\phi}(\vec{y}) \normord{ \phi(\vec{x}) \phi(\vec{y}) } \\
&\quad+ k_{c,+}(\vec{x}) k_{s,\phi}(\vec{y}) \normord{ \pi(\vec{x}) \phi(\vec{y}) } + k_{s,\phi}(\vec{x}) k_{c,+}(\vec{y}) \normord{ \phi(\vec{x}) \pi(\vec{y}) } \\
&\quad+ \left[ k_{c,+}(\vec{x}) k_{c,+}(\vec{y}) - 1 \right] \expect{ \pi(\vec{x}) \pi(\vec{y}) } + k_{s,\phi}(\vec{x}) k_{s,\phi}(\vec{y}) \expect{ \phi(\vec{x}) \phi(\vec{y}) } \\
&\quad+ k_{c,+}(\vec{x}) k_{s,\phi}(\vec{y}) \expect{ \pi(\vec{x}) \phi(\vec{y}) } + k_{s,\phi}(\vec{x}) k_{c,+}(\vec{y}) \expect{ \phi(\vec{x}) \pi(\vec{y}) } \eqend{,}
\end{split}
\end{equations}
where we set $\expect{ \cdot } \defby \left( \Omega, \cdot \, \Omega \right)$. Note also the special case~\eqref{eq:timeslicesqueezing}, where we have $k_0 = r$ and $k_\phi = k_\pi = 0$, which gives $k_{c,\pm}(\vec{x}) = \exp(\pm r(\vec{x}))$ and $k_{s,\phi}(\vec{x}) = k_{s,\pi}(\vec{x}) = 0$.

From \eqref{eq:nordprod1} and \eqref{eq:nordprod2} we obtain $S(k)^\dagger \normord{\phi^2}(\vec{x}) S(k)$ and $S(k)^\dagger \normord{\pi^2}(\vec{x})S(k)$ by taking the limit $\vec{y} \to \vec{x}$. Taking expectation values and using that $\expect{ \{ \phi(\vec{x}), \pi(\vec{x}) \} } = 0$, we obtain
\begin{equations}
\expect{S(k)^\dagger \normord{\phi^2}(\vec{x}) S(k)} = \left[ k_{c,-}(\vec{x})^2 - 1 \right] \expect{ \phi(\vec{x})^2 } + k_{s,\pi}(\vec{x})^2 \expect{ \pi(\vec{x})^2 } \eqend{,} \\
\expect{S(k)^\dagger \normord{ \pi^2}(\vec{x})  S(k)} = \left[ k_{c,+}(\vec{x})^2   - 1 \right] \expect{ \pi(\vec{x})^2 } + k_{s,\phi}(\vec{x})^2 \expect{ \phi(\vec{x})^2 } \eqend{.}
\end{equations}
It is also straightforward to obtain $\sum_{i=1}^D S(k)^\dagger \normord{(\partial_i \phi)^2}(\vec{x}) S(k)$ and its expectation value by applying suitable derivatives in $\vec{x}$ and $\vec{y}$ before taking the limit $\vec{y} \to \vec{x}$. For $i \in \{ 1,\ldots, D\}$, this yields
\begin{equations}
\begin{split}
    \expect{S(k)^\dagger \normord{(\partial_i \phi)^2}(\vec{x}) S(k)}
    &= (\partial_i k_{c,-}(\vec{x}))^2 \expect{\phi(\vec{x})^2} + [k_{c,-}(\vec{x})^2-1 ]\expect{(\partial_i \phi(\vec{x}))^2} \\
    & + (\partial_{i} k_{s,\pi}(\vec{x}))^2 \expect{\pi(\vec{x})^2} + k_{s,\pi}(\vec{x})^2 \expect{(\partial_i \pi(\vec{x}))^2} \\
    & + \lim_{\vec{y}\to\vec{x}} [k_{c,-}(\vec{x}) \partial_i k_{c,-} (\vec{y})- \partial_i k_{c,-}(\vec{x}) k_{c,-}(\vec{y})] \expect{ \partial_i \phi (\vec{x}) \phi(\vec{y})} \\
    & + \lim_{\vec{y}\to\vec{x}}  [k_{s,\pi}(\vec{x}) \partial_i k_{s,\pi} (\vec{y})- \partial_i k_{s,\pi}(\vec{x}) k_{s,\pi}(\vec{y})] \expect{\partial_i \pi(\vec{x}) \pi(\vec{y})}.
    \end{split}
\end{equations}

Finally, using the relation $\expect{ \sum_{i=1}^D \bigl[ \partial_i \phi(\vec{x}) \bigr]^2 } = \expect{ \pi(\vec{x})^2 } - m^2 \expect{ \phi(\vec{x})^2 }$ that follows from the Klein--Gordon equation for $\phi$ and the fact that the correlation functions in the Minkowski vacuum state $\ket{\Omega}$ only depend on the differences of the coordinates, we obtain
\begin{equation}
\begin{split}
\expect{T_S^{00}(0,\vec{x})} &= \frac{1}{2} \biggl[ F_1(\vec{x}) \expect{ \phi(\vec{x})^2 } + F_2(\vec{x}) \expect{ \pi(\vec{x})^2 } + F_3(\vec{x}) \bigg\langle \sum_{i=1}^D \bigl[ \partial_i \pi(\vec{x}) \bigr]^2 \bigg\rangle \\
&+  \lim_{\vec{y}\to\vec{x}} \sum_{i=1}^D F_{4,i}(\vec{x},\vec{y}) \bigg\langle \sum_{i=1}^D \bigl[ \partial_i \phi(\vec{x}) \phi(\vec{y}) \bigr] \bigg\rangle + \lim_{\vec{y}\to\vec{x}} \sum_{i=1}^D F_{5,i} (\vec{x},\vec{y}) \bigg\langle \bigl[ \partial_i \pi(\vec{x}) \pi(\vec{y}) \bigr] \bigg\rangle \biggr]
\end{split}
\end{equation}
with
\begin{equations}
F_1(\vec{x}) &\defby k_{s,\phi}(\vec{x})^2 + \sum_{i=1}^D (\partial_i k_{c,-}(\vec{x}))^2 \eqend{,} \\
F_2(\vec{x}) &\defby k_{c,+}(\vec{x})^2 + k_{c,-}(\vec{x})^2 - 2 + m^2 k_{s,\pi}(\vec{x})^2 + \sum_{i=1}^D(\partial_i k_{s,\pi}(\vec{x}))^2\eqend{,} \\
F_3(\vec{x}) &\defby k_{s,\pi}(\vec{x})^2 \eqend{,} \\
F_{4,i}(\vec{x},\vec{y}) & \defby k_{c,-}(\vec{x}) \partial_i k_{c,-}(\vec{y}) - \partial_i k_{c,-}(\vec{x})  k_{c,-}(\vec{y}) \eqend{,} \\
F_{5,i} (\vec{x},\vec{y}) & \defby  k_{s,\pi}(\vec{x}) \partial_i k_{s,\pi}(\vec{y}) - \partial_i k_{s,\pi}(\vec{x})  k_{s,\pi}(\vec{y}) \eqend{.}
\end{equations}

Since all appearing vacuum expectation values are divergent and independent of each other, our computation shows that the energy density is divergent unless all the coefficients $F_1,F_2,F_3$ and the respective limits dependent on $F_{4,i}$ and $F_{5,i}$ are vanishing. This first implies $k_{s,\pi}(\vec{x}) = 0$ and $k_{c,+}^2 + k_{c,-}^2 = 2$. Since our variables obey the relation $k_{c,+} k_{c,-} + k_{s,\phi} k_{s,\pi} = 1$, we may then infer $k_{c,+} = k_{c,-} \in \{ \pm 1\}$ and lastly $k_{s,\phi} = 0$. In turn, this implies $S(k) \Omega = \mathe^{\mathi c} \, \Omega$ with $c \in \mathbb{R}$, i.e., \emph{no squeezing at all}.

\section{Norm estimates for local squeezing}
\label{app:normestimates}

We define the norms
\begin{equation}
\norm{ f }_p \defby \left[ \int \frac{1}{2 \omega_\vec{k}} \abs{ \tilde{f}(\omega_\vec{k},\vec{k}) }^p \frac{\total^D \vec{k}}{(2\pi)^D} \right]^\frac{1}{p} \eqend{,} \quad 1 \leq p \leq \infty \eqend{,}
\end{equation}
which are just the $L^p$ norms of the Fourier transform of $f$ restricted to the mass shell, with respect to the measure
\begin{equation}
\label{eq:hilbert_1p_measure}
\total \mu(\vec{k}) = \frac{1}{2 \omega_\vec{k}} \frac{\total^D \vec{k}}{(2 \pi)^D}
\end{equation}
and with the usual modification for $p = \infty$. In particular, for $p = 2$ we have
\begin{equation}
\norm{ f }_2^2 = \norm{f}_\omega^2
\end{equation}
and recover the norm of the single-particle Hilbert space $\mathcal{H}$.

In order to prove \Cref{lem:thdef}, for $f, h \in \mathcal{S}_\mathbb{R}(\mathbb{R}^d)$ we need to show that
\begin{equation}
\norm{ t_h f }_2 \leq c(h) \norm{f}_2 < \infty
\end{equation}
for a suitable finite constant $c(h)$. To do so, we employ interpolation theory, in particular, the Riesz--Thorin theorem\footnote{We use the formulation from \cite{SS11}.}:
\begin{theorem}[Riesz--Thorin]
\label{thm:riesz_thorin}
Let $X$, $Y$ be $\sigma$-finite measure spaces and let $1 \leq r,s \leq\infty$. Let $T \colon L^r(X) + L^s(X) \rightarrow L^r(Y) + L^s(Y)$ be a linear operator obeying for $p = r, s$ and some nonnegative constants $B_p$ the bounds $\norm{ T f }_{L^p(Y)} \leq B_p \norm{ f }_{L^p(X)}$ for all $f \in L^p(X)$. Then we have
\begin{equation}
\norm{ T f }_{L^{p_\theta}(Y)} \leq B_\theta \norm{ f }_{L^{p_\theta}(X)}
\end{equation}
for all $0 \leq \theta \leq 1$ and $f \in L^{p_\theta}(X)$, where $1/p_\theta \defby (1-\theta)/r + \theta/s$ and $B_\theta \defby B_r^{1-\theta} B_s^\theta $.
\end{theorem}
Note that it is clearly sufficient to establish the bounds $\norm{ T f }_{L^p(Y)} \leq B_p \norm{ f }_{L^p(X)}$ for all $f$ in some dense subspace of $L^p(X)$. Choosing $X = Y = (\mathbb{R}^D, \total \mu)$, $T = t_h$, $r = 1$, $s = \infty$ and $\theta = \frac{1}{2}$ so that $p_\theta = 2$, we thus need to show that for $p = 1$ and $p = \infty$ there are nonnegative constants $B_p$ such that
\begin{equation}
\norm{ t_h f }_p \leq B_p \norm{ f }_p \eqend{,}
\end{equation}
which by the Riesz--Thorin theorem then implies
\begin{equation}
\norm{ t_h f }_2 \leq \sqrt{ B_1 B_\infty } \norm{ f }_2 \eqend{.}
\end{equation}

Using the definition~\eqref{eq:th1} of $t_h$ for arguments from $\mathcal{S}_\mathbb{R}(\mathbb{R}^d)$ and the Fourier transform of $\Delta$~\eqref{eq:Delta}, we first obtain the Fourier transform of $t_h f$ in case that $f$ is real-valued:
\begin{equation}
\label{eq:thid}
(\widetilde{t_h f})(\omega_\vec{p}, \vec{p}) = - \mathi \int \frac{1}{2 \omega_\vec{k}} \left[ \tilde{h}(\omega_\vec{p} - \omega_\vec{k}, \vec{p}-\vec{k}) \tilde{f}(\omega_\vec{k}, \vec{k}) - \tilde{h}(\omega_\vec{p} + \omega_\vec{k}, \vec{p}-\vec{k}) \tilde{f}(-\omega_\vec{k}, \vec{k}) \right] \frac{\total^D \vec{k}}{(2 \pi)^D} \eqend{.}
\end{equation}
For $p = 1$, we then compute
\begin{splitequation}
\norm{ t_h f }_1 &= \int \frac{1}{2 \omega_\vec{p}} \abs{ (\widetilde{t_h f})(\omega_\vec{p}, \vec{p}) } \frac{\total^D \vec{p}}{(2\pi)^D} \\
&\leq \iint \frac{1}{4 \omega_\vec{p} \omega_\vec{k}} \left[ \abs{ \tilde{h}(\omega_\vec{p} - \omega_\vec{k}, \vec{p}-\vec{k}) } \abs{ \tilde{f}(\omega_\vec{k}, \vec{k}) } + \abs{ \tilde{h}(\omega_\vec{p} + \omega_\vec{k}, \vec{p}-\vec{k}) } \abs{ \tilde{f}(-\omega_\vec{k}, \vec{k}) } \right] \frac{\total^D \vec{k}}{(2 \pi)^D} \frac{\total^D \vec{p}}{(2\pi)^D} \\
&= \iint \frac{1}{4 \omega_{\vec{p}+\vec{k}} \omega_\vec{k}} \left[ \abs{ \tilde{h}(\omega_{\vec{p}+\vec{k}} - \omega_\vec{k}, \vec{p}) } \abs{ \tilde{f}(\omega_\vec{k}, \vec{k}) } + \abs{ \tilde{h}(\omega_{\vec{p}+\vec{k}} + \omega_\vec{k}, \vec{p}) } \abs{ \tilde{f}(-\omega_\vec{k}, \vec{k}) } \right] \frac{\total^D \vec{p}}{(2\pi)^D} \frac{\total^D \vec{k}}{(2 \pi)^D} \eqend{,}
\end{splitequation}
where we employed Tonelli's theorem to interchange the integrations and afterwards shifted $\vec{p} \to \vec{p} + \vec{k}$. We now define
\begin{equation}
\label{eq:ch_def}
c(h) \defby \sup_{\vec{p},\vec{k}} \left( \frac{\omega_\vec{p}^{D+1}}{2m^2} \abs{ \tilde{h}(\pm \omega_{\vec{p}+\vec{k}} \pm \omega_{\vec{k}}, \vec{p}) } \right) < \infty \eqend{,}
\end{equation}
where the two signs are independent of each other and which is finite since $h$, and thus also $\tilde{h}$, is a Schwartz function. We thus have the estimate
\begin{equation}
\label{eq:h_in_c_bound}
\abs{ \tilde{h}(\omega_{\vec{p}+\vec{k}} \pm \omega_{\vec{k}}, \vec{p}) } \leq 2c(h) \frac{m^2}{\omega_\vec{p}^{D+1}} \eqend{,}
\end{equation}
and using also $\omega_{\vec{p}+\vec{k}} \geq m$ and that
\begin{equation}
\label{eq:did}
\int \frac{2 m}{\omega_\vec{p}^{D+1}} \frac{\total^D \vec{p}}{(2\pi)^D} = \frac{1}{2^{D-1} \pi^\frac{D-1}{2} \Gamma\left( \frac{D+1}{2} \right)} \leq 1 \eqend{,}
\end{equation}
it follows that
\begin{splitequation}
\norm{ t_h f }_1 &\leq c(h) \iint \frac{m}{2 \omega_\vec{k} \omega_\vec{p}^{D+1}} \left[ \abs{ \tilde{f}(\omega_\vec{k}, \vec{k}) } + \abs{ \tilde{f}(-\omega_\vec{k}, \vec{k}) } \right] \frac{\total^D \vec{p}}{(2\pi)^D} \frac{\total^D \vec{k}}{(2 \pi)^D} \\
&\leq \frac{c(h)}{2} \int \frac{1}{2 \omega_\vec{k}} \left[ \abs{ \tilde{f}(\omega_\vec{k}, \vec{k}) } + \abs{ \tilde{f}(-\omega_\vec{k}, \vec{k}) } \right] \frac{\total^D \vec{k}}{(2 \pi)^D} \eqend{.}
\end{splitequation}
For real $f$, we have $\tilde{f}(- \omega_\vec{k}, \vec{k}) = \tilde{f}(\omega_\vec{k}, - \vec{k})$, such that changing $\vec{k} \to - \vec{k}$ in the second term results in
\begin{equation}
\norm{ t_h f }_1 \leq c(h) \norm{ f }_1 \eqend{,}
\end{equation}
which shows that $B_1 = c(h)$.

For $p = \infty$, we obtain instead
\begin{splitequation}
\norm{ t_h f }_\infty &= \esssup_\vec{p} \abs{ (\widetilde{t_h f})(\omega_\vec{p},\vec{p}) } \\
&\leq \esssup_\vec{p} \left[ \int \abs{ \tilde{h}(\omega_\vec{p} - \omega_\vec{k}, \vec{p}-\vec{k}) } \frac{\total^D \vec{k}}{(2 \pi)^D} \right] \esssup_\vec{k} \left[ \frac{1}{2 \omega_\vec{k}} \abs{ \tilde{f}(\omega_\vec{k}, \vec{k}) } \right] \\
&\quad+ \esssup_\vec{p} \left[ \int \abs{ \tilde{h}(\omega_\vec{p} + \omega_\vec{k}, \vec{p}-\vec{k}) } \frac{\total^D \vec{k}}{(2 \pi)^D} \right] \esssup_\vec{k} \left[ \frac{1}{2 \omega_\vec{k}} \abs{ \tilde{f}(- \omega_\vec{k}, \vec{k}) } \right] \eqend{,}
\end{splitequation}
and then perform the change of variable $\vec{k} \to \vec{p} - \vec{k}$ in the integrals. This gives
\begin{splitequation}
\norm{ t_h f }_\infty &\leq \esssup_\vec{p} \left[ \int \abs{ \tilde{h}(\omega_\vec{p} - \omega_{\vec{p} - \vec{k}}, \vec{k}) } \frac{\total^D \vec{k}}{(2 \pi)^D} \right] \esssup_\vec{k} \left[ \frac{1}{2 \omega_\vec{k}} \abs{ \tilde{f}(\omega_\vec{k}, \vec{k}) } \right] \\
&\quad+ \esssup_\vec{p} \left[ \int \abs{ \tilde{h}(\omega_\vec{p} + \omega_{\vec{p} - \vec{k}}, \vec{k}) } \frac{\total^D \vec{k}}{(2 \pi)^D} \right] \esssup_\vec{k} \left[ \frac{1}{2 \omega_\vec{k}} \abs{ \tilde{f}(- \omega_\vec{k}, \vec{k}) } \right] \\
&\leq m c(h) \int \frac{1}{\omega_\vec{k}^{D+1}} \frac{\total^D \vec{k}}{(2 \pi)^D} \, \esssup_\vec{k} \abs{ \tilde{f}(\omega_\vec{k}, \vec{k}) } \\
&\quad+ m c(h) \int \frac{1}{\omega_\vec{k}^{D+1}} \frac{\total^D \vec{k}}{(2 \pi)^D} \, \esssup_\vec{k} \abs{ \tilde{f}(- \omega_\vec{k}, \vec{k}) } \\
&\leq \frac{c(h)}{2} \left[ \esssup_\vec{k} \abs{ \tilde{f}(\omega_\vec{k}, \vec{k}) } + \esssup_\vec{k} \abs{ \tilde{f}(- \omega_\vec{k}, \vec{k}) } \right]
\end{splitequation}
with the same estimate on $\tilde{h}$ as before. For real-valued $f$, we have $\abs{ \tilde{f}(- \omega_\vec{k}, \vec{k}) } = \abs{ \tilde{f}(\omega_\vec{k}, - \vec{k}) }$, which has the same essential supremum as $\abs{ \tilde{f}(\omega_\vec{k}, \vec{k}) }$, and thus the estimate
\begin{equation}
\norm{ t_h f }_\infty \leq c(h) \norm{ f }_\infty \eqend{,}
\end{equation}
which shows that also $B_\infty = c(h)$.

As explained before, we thus obtain
\begin{equation}
\label{eq:thf_norm_2_estimate}
\norm{ t_h f }_2 \leq c(h) \norm{f}_2 \eqend{,}
\end{equation}
and in fact the same estimate for all $p \in [1,\infty]$.

\section{Estimates for Wick powers}
\label{app:wickestimates}

In order to prove \Cref{lem:wicksadj}, we first consider the action of a single combination $\Phi(f) + \normord{ \Phi^2 }(h)$ on an $n$-particle vector $\psi^{(n)} \in \mathcal{F}_0$, where $f, h \in \mathcal{S}_\mathbb{R}(\mathbb{R}^d)$.

We use the conventions of~\cite[Ch. 5.2]{bratellirobinson2} or~\cite[Ch.~X.7]{reedsimon2}. In particular, for real-valued $f$ the field operator is given by
\begin{equation}
\Phi(f) = \frac{1}{\sqrt{2}} \overline{ \left( a(f) + a^\dagger(f) \right) }
\end{equation}
in terms of the annihilation operator $a(f)$ and the creation operator $a^\dagger(f)$. The annihilation and creation operators correspond to operator-valued distributions $a(\vec{p})$ and $a^\dagger(\vec{p})$ such that
\begin{equation}
a(f) = \int \overline{\tilde{f}(\omega_\vec{p}, \vec{p})} \, a(\vec{p}) \total \mu(\vec{p}) \eqend{,} \quad a^\dagger(f) = \int \tilde{f}(\omega_\vec{p}, \vec{p}) \, a^\dagger(\vec{p}) \total \mu(\vec{p}) \eqend{,}
\end{equation}
with the measure $\total \mu(\vec{p})$ given by \eqref{eq:hilbert_1p_measure}. The elements of the bosonic $n$-particle space $\mathcal{H}^{(n)}_\mathrm{s}$ are vectors
\begin{equation}
\psi^{(n)}(g) = \frac{1}{\sqrt{n!}} \int\dotsi\int g(\vec{p}_1,\ldots,\vec{p}_n) a^\dagger(\vec{p}_1) \cdots a^\dagger(\vec{p}_n) \Omega \total \mu(\vec{p}_1) \cdots \total \mu(\vec{p}_n) \eqend{,}
\end{equation}
whose norm is given by
\begin{equation}
\norm{ \psi^{(n)}(g) }^2 = \int\dotsi\int \abs{ g(\vec{p}_1,\ldots,\vec{p}_n) }^2 \total \mu(\vec{p}_1) \cdots \total \mu(\vec{p}_n) \eqend{.}
\end{equation}

Acting with the field operator $\Phi(f)$ on a finite-particle vector $\psi^{(n)}(g)$, we obtain
\begin{splitequation}
\Phi(f) \psi^{(n)}(g) &= \frac{1}{\sqrt{2 n!}} \sum_{k=1}^n \int\dotsi\int \tilde{f}(\omega_{\vec{p}_k}, \vec{p}_k) g(\vec{p}_1,\ldots,\vec{p}_n) \\
&\qquad\times a^\dagger(\vec{p}_1) \cdots a^\dagger(\vec{p}_{k-1}) a^\dagger(\vec{p}_{k+1}) \cdots a^\dagger(\vec{p}_n) \Omega \total \mu(\vec{p}_1) \cdots \total \mu(\vec{p}_n) \\
&\quad+ \frac{1}{\sqrt{2 n!}} \int\dotsi\int \tilde{f}(\omega_\vec{p}, \vec{p}) g(\vec{p}_1,\ldots,\vec{p}_n) \\
&\qquad\times a^\dagger(\vec{p}) a^\dagger(\vec{p}_1) \cdots a^\dagger(\vec{p}_n) \Omega \total \mu(\vec{p}_1) \cdots \total \mu(\vec{p}_n) \total \mu(\vec{p}) \eqend{,}
\end{splitequation}
where we repeatedly commuted the annihilation operator to the right, using that
\begin{equation}
\int f(\vec{p}) [ a(\vec{p}), a^\dagger(\vec{q}) ] \total \mu(\vec{p}) = f(\vec{q}) \eqend{.}
\end{equation}
Because $g$ is symmetric in its arguments, the sum in the first term gives $n$ times the same result (say with $k = 1$), and we see that we obtain a combination of an $(n-1)$-particle vector and an $(n+1)$-particle vector.

For the normal-ordered square of the field, we compute
\begin{splitequation}
\normord{ \Phi^2 }(h) &= \frac{1}{2} \iint \biggl[ a(\vec{p}) a(\vec{q}) \tilde{h}(-\omega_\vec{p}-\omega_\vec{q},-\vec{p}-\vec{q}) + 2 a^\dagger(\vec{p}) a(\vec{q}) \tilde{h}(\omega_\vec{p}-\omega_\vec{q}, \vec{p}-\vec{q}) \\
&\qquad\qquad+ a^\dagger(\vec{p}) a^\dagger(\vec{q}) \tilde{h}(\omega_\vec{p}+\omega_\vec{q},\vec{p}+\vec{q}) \biggr] \total \mu(\vec{p}) \total \mu(\vec{q}) \eqend{,}
\end{splitequation}
and thus acting on a finite-particle vector $\psi^{(n)}(g)$ we obtain
\begin{splitequation}
\normord{ \Phi^2 }(h) \psi^{(n)}(g) &= \frac{1}{\sqrt{n!}} \sum_{k=1}^{n-1} \sum_{\ell=k+1}^n \int\dotsi\int \tilde{h}(-\omega_{\vec{p}_\ell}-\omega_{\vec{p}_k},-\vec{p}_\ell-\vec{p}_k) g(\vec{p}_1,\ldots,\vec{p}_n) \\
&\qquad\times a^\dagger(\vec{p}_1) \cdots a^\dagger(\vec{p}_{k-1}) a^\dagger(\vec{p}_{k+1}) \cdots a^\dagger(\vec{p}_{\ell-1}) a^\dagger(\vec{p}_{\ell+1}) \cdots a^\dagger(\vec{p}_n) \Omega \total \mu(\vec{p}_1) \cdots \total \mu(\vec{p}_n) \\
&\quad+ \frac{1}{\sqrt{n!}} \sum_{k=1}^n \int\dotsi\int \tilde{h}(\omega_\vec{p}-\omega_{\vec{p}_k},\vec{p}-\vec{p}_k) g(\vec{p}_1,\ldots,\vec{p}_n) \\
&\qquad\times a^\dagger(\vec{p}) a^\dagger(\vec{p}_1) \cdots a^\dagger(\vec{p}_{k-1}) a^\dagger(\vec{p}_{k+1}) \cdots a^\dagger(\vec{p}_n) \Omega \total \mu(\vec{p}_1) \cdots \total \mu(\vec{p}_n) \total \mu(\vec{p}) \\
&\quad+ \frac{1}{2 \sqrt{n!}} \int\dotsi\int \tilde{h}(\omega_\vec{p}+\omega_\vec{q},\vec{p}+\vec{q}) g(\vec{p}_1,\ldots,\vec{p}_n) \\
&\qquad\times a^\dagger(\vec{p}) a^\dagger(\vec{q}) a^\dagger(\vec{p}_1) \cdots a^\dagger(\vec{p}_n) \Omega \total \mu(\vec{p}_1) \cdots \total \mu(\vec{p}_n) \total \mu(\vec{p}) \total \mu(\vec{q}) \eqend{,}
\end{splitequation}
where we again repeatedly commuted the annihilation operators to the right and in addition in the first term used that they commute among themselves. Because $g$ is symmetric in its arguments, the sums in the first term give $n(n-1)/2$ times the same result (say with $k = 1$, $\ell = 2$), and the sum in the second term gives $n$ times the same result, and we obtain a combination of an $(n-2)$-particle vector, an $n$-particle vector, and an $(n+2)$-particle vector.

Acting with the linear combination $\Phi(f) + \normord{ \Phi^2 }(h)$ on $\psi^{(n)}(g)$, we thus obtain a sum of $k$-particle vectors with $k \in \{ n-2, n-1, n, n+1, n+2 \}$, such that the square of the norm of the left-hand side is given by the sum of the square of the norms of each of these. It follows that
\begin{splitequation}
&\norm{ \left[ \Phi(f) + \normord{ \Phi^2 }(h) \right] \psi^{(n)}(g) }^2 \\
&= \frac{n}{2} \int\dotsi\int \abs{ \int \tilde{f}(\omega_{\vec{p}_1}, \vec{p}_1) g(\vec{p}_1,\ldots,\vec{p}_n) \total \mu(\vec{p}_1) }^2 \total \mu(\vec{p}_2) \cdots \total \mu(\vec{p}_n) \\
&\quad+ \frac{n+1}{2} \int\dotsi\int \abs{ \tilde{f}(\omega_\vec{p}, \vec{p}) g(\vec{p}_1,\ldots,\vec{p}_n) }^2 \total \mu(\vec{p}_1) \cdots \total \mu(\vec{p}_n) \total \mu(\vec{p}) \\
&\quad+ \frac{n (n-1)}{4} \int\dotsi\int \abs{ \iint \tilde{h}(-\omega_{\vec{p}_1}-\omega_{\vec{p}_2},-\vec{p}_1-\vec{p}_2) g(\vec{p}_1,\ldots,\vec{p}_n) \total \mu(\vec{p}_1) \total \mu(\vec{p}_2) }^2 \\
&\qquad\qquad\times \total \mu(\vec{p}_3) \cdots \total \mu(\vec{p}_n) \\
&\quad+ \frac{(n+2) (n+1)}{4} \int\dotsi\int \abs{ \tilde{h}(\omega_\vec{p}+\omega_\vec{q},\vec{p}+\vec{q}) g(\vec{p}_1,\ldots,\vec{p}_n) }^2 \\
&\qquad\qquad\times \total \mu(\vec{p}_1) \cdots \total \mu(\vec{p}_n) \total \mu(\vec{p}) \total \mu(\vec{q}) \\
&\quad+ n^2 \int\dotsi\int \abs{ \int \tilde{h}(\omega_\vec{p}-\omega_{\vec{p}_1},\vec{p}-\vec{p}_1) g(\vec{p}_1,\ldots,\vec{p}_n) \total \mu(\vec{p}_1) }^2 \total \mu(\vec{p}) \total \mu(\vec{p}_2) \cdots \total \mu(\vec{p}_n) \eqend{.}
\end{splitequation}
For the first term, we use the Cauchy--Schwarz inequality to obtain
\begin{equation}
\abs{ \int \tilde{f}(\omega_{\vec{p}_1}, \vec{p}_1) g(\vec{p}_1,\ldots,\vec{p}_n) \total \mu(\vec{p}_1) }^2 \leq \int \abs{ \tilde{f}(\omega_{\vec{p}_1}, \vec{p}_1) }^2 \total \mu(\vec{p}_1) \int \abs{ g(\vec{p}_1,\ldots,\vec{p}_n) }^2 \total \mu(\vec{p}_1) \eqend{,}
\end{equation}
and analogously we obtain bounds for all other terms except the last one. In this term, the integrand does not decay for $\vec{p}_1 \approx \vec{p}$, and we need to be more careful to obtain a suitable bound. Ref. \cite{LS65} uses a result of Dixmier and shows boundedness of an associated bilinear form to estimate this term, but a much simpler bound can be obtained using the Riesz--Thorin theorem.

Consider thus the operator
\begin{equation}
S_h\colon g(\vec{p}_1,\ldots,\vec{p}_n) \mapsto \int \tilde{h}(\omega_{\vec{p}_1}-\omega_\vec{p},\vec{p}_1-\vec{p}) g(\vec{p},\vec{p}_2,\ldots,\vec{p}_n) \total \mu(\vec{p}) \eqend{.}
\end{equation}
Choosing in \Cref{thm:riesz_thorin} $X = Y = (\mathbb{R}^{n D}, \total \mu^{\otimes n})$, $T = S_h$, $r = 1$, $s = \infty$ and $\theta = \frac{1}{2}$ so that $p_\theta = 2$ and noting that it is sufficient to establish the bounds for a dense subspace, we thus need to show that for $p = 1$ and $p = \infty$ there are nonnegative constants $B_p$ such that for all $g,h \in S_\mathbb{R}(\mathbb{R}^d)$,
\begin{equation}
\norm{ S_h g }_p \leq B_p \norm{ g }_p \eqend{,}
\end{equation}
which by the Riesz--Thorin theorem then implies
\begin{equation}
\norm{ S_h g }_2 \leq \sqrt{ B_1 B_\infty } \norm{ g }_2 \eqend{,}
\end{equation}
which is the estimate that we need.

For the 1-norm, we obtain
\begin{splitequation}
\norm{ S_h g }_1 &= \int\dotsi\int \abs{ (S_h g)(\vec{p}_1,\ldots,\vec{p}_n) } \total \mu(\vec{p}_1) \cdots \total \mu(\vec{p}_n) \\
&\leq \int\dotsi\int \abs{ \tilde{h}(\omega_{\vec{p}_1}-\omega_\vec{p},\vec{p}_1-\vec{p}) } \abs{ g(\vec{p},\vec{p}_2,\ldots,\vec{p}_n) } \total \mu(\vec{p}) \total \mu(\vec{p}_1) \cdots \total \mu(\vec{p}_n) \\
&\leq \frac{1}{2 m} \int\dotsi\int \abs{ \tilde{h}(\omega_{\vec{p}_1+\vec{p}}-\omega_\vec{p},\vec{p}_1)} \abs{ g(\vec{p},\vec{p}_2,\ldots,\vec{p}_n) } \frac{\total^D \vec{p}_1}{(2\pi)^D} \total \mu(\vec{p}) \total \mu(\vec{p}_2) \cdots \total \mu(\vec{p}_n) \\
&\leq c(h) \int \frac{m}{\omega_{\vec{p}_1}^{D+1}} \frac{\total^D \vec{p}_1}{(2\pi)^D} \int\dotsi\int \abs{g(\vec{p},\vec{p}_2,\ldots,\vec{p}_n) } \total \mu(\vec{p}) \total \mu(\vec{p}_2) \cdots \total \mu(\vec{p}_n) \\
&= \frac{c(h)}{2} \norm{ g }_1 \eqend{,}
\end{splitequation}
where we shifted $\vec{p}_1 \to \vec{p}_1 + \vec{p}$, used as before that $\omega_\vec{p} \geq m$, and employed the bound~\eqref{eq:h_in_c_bound} for $\tilde{h}$ and the integral~\eqref{eq:did}, such that $B_1 = c(h)/2$. For the infinity norm, the analogous computation establishes
\begin{splitequation}
\norm{ S_h g }_\infty &= \esssup_{\vec{p}_1,\ldots,\vec{p}_n} \abs{ (S_h g)(\vec{p}_1,\ldots,\vec{p}_n) } \\
&\leq \esssup_{\vec{p}_1,\ldots,\vec{p}_n} \int \abs{ \tilde{h}(\omega_{\vec{p}_1}-\omega_\vec{p},\vec{p}_1-\vec{p})} \abs{ g(\vec{p},\vec{p}_2,\ldots,\vec{p}_n) } \total \mu(\vec{p}) \\
&\leq \frac{1}{2 m} \esssup_{\vec{p}_1,\ldots,\vec{p}_n} \int \abs{ \tilde{h}(\omega_{\vec{p}_1}-\omega_{\vec{p}-\vec{p}_1},\vec{p})} \abs{ g(\vec{p}_1-\vec{p},\vec{p}_2,\ldots,\vec{p}_n) } \frac{\total^{D} \vec{p}}{(2\pi)^{D}} \\
&\leq c(h) \int \frac{m}{\omega_\vec{p}^{D+1}} \esssup_{\vec{p}_1,\ldots,\vec{p}_n} \abs{ g(\vec{p}_1-\vec{p},\vec{p}_2,\ldots,\vec{p}_n) } \frac{\total^{D} \vec{p}}{(2\pi)^{D}} \leq \frac{c(h)}{2} \norm{g }_\infty \eqend{,}
\end{splitequation}
where we changed variables to $\vec{p} \to \vec{p}_1 - \vec{p}$, and then used the same bound~\eqref{eq:h_in_c_bound} on $\tilde{h}$ and integral~\eqref{eq:did} as before, and thus also $B_\infty = c(h)/2$. By the Riesz--Thorin interpolation theorem, it thus follows that
\begin{splitequation}
&\int\dotsi\int \abs{ \int \tilde{h}(\omega_\vec{p}-\omega_{\vec{p}_1},\vec{p}-\vec{p}_1) g(\vec{p}_1,\ldots,\vec{p}_n) \total \mu(\vec{p}_1) }^2 \total \mu(\vec{p}) \total \mu(\vec{p}_2) \cdots \total \mu(\vec{p}_n) \\
&= \norm{ S_h g }_2^2 \leq B_1 B_\infty \norm{ g }_2^2 = \frac{[ c(h) ]^2}{4} \int\dotsi\int \abs{ g(\vec{p}_1,\ldots,\vec{p}_n) }^2 \total \mu(\vec{p}_1) \total \mu(\vec{p}_2) \cdots \total \mu(\vec{p}_n) \eqend{.}
\end{splitequation}

We now define
\begin{splitequation}
\label{eq:kalphabeta_def}
K(f,h) &\defby \max\biggl[ \left( \int \abs{ \tilde{f}(\omega_\vec{p}, \vec{p}) }^2 \total \mu(\vec{p}) \right)^\frac{1}{2}, 2 c(h), \\
&\qquad\qquad \left( \iint \abs{ \tilde{h}(\pm (\omega_\vec{p}+\omega_\vec{q}), \vec{p}+\vec{q}) }^2 \total \mu(\vec{p}) \total \mu(\vec{q}) \right)^\frac{1}{2} \biggr] \eqend{,}
\end{splitequation}
where the first integral is finite since $f \in L^2(\mathbb{R}^D, \total \mu)$ and the last integral is finite since $h$, and thus also $\tilde{h}$, is a Schwartz function and thus also satisfies the bound
\begin{equation}
\abs{ \tilde{h}(\pm (\omega_\vec{p}+\omega_\vec{q}), \vec{p}+\vec{q}) } \leq \frac{c'}{(\omega_\vec{p} + \omega_\vec{q})^{D+1}} \leq \frac{c'}{\omega_\vec{p}^\frac{D+1}{2} \omega_\vec{q}^\frac{D+1}{2}}
\end{equation}
for some constant $c'$, and the integral over $\vec{p}$ and $\vec{q}$ of the square of the right-hand side is finite. Taking all together, we obtain the bound
\begin{splitequation}
&\norm{ \left[ \Phi(f) + \normord{ \Phi^2 }(h) \right] \psi^{(n)}(g) }^2 \\
&\quad\leq \frac{(3n+4)^2}{16} [ K(f,h) ]^2 \int\dotsi\int \abs{ g(\vec{p}_1,\ldots,\vec{p}_n) }^2 \total \mu(\vec{p}_1) \cdots \total \mu(\vec{p}_n) \\
&\quad\leq (n+1)^2 [ K(f,h) ]^2 \norm{ \psi^{(n)}(g) }^2 \eqend{,}
\end{splitequation}
and thus
\begin{equation}
\label{eq:fixedparticlebound_1}
\norm{ \left[ \Phi(f) + \normord{ \Phi^2 }(h) \right] \psi^{(n)}(g) } \leq (n+1) K(f, h) \norm{ \psi^{(n)}(g) } \eqend{.}
\end{equation}

Applying $\left[ \Phi(f) + \normord{ \Phi^2 }(h) \right]$ again, we then need to take into account that we now have 5 vectors with at most $n+2$ particles, and thus obtain the bound
\begin{splitequation}
&\norm{ \left[ \Phi(f_1) + \normord{ \Phi^2 }(h_1) \right] \left[ \Phi(f_2) + \normord{ \Phi^2 }(h_2) \right] \psi^{(n)}(g) } \\
&\quad\leq 5 (n+3) K(f_1, h_1) \norm{ \left[ \Phi(f_2) + \normord{ \Phi^2 }(h_2) \right] \psi^{(n)}(g) } \\
&\quad\leq 5^2 (n+1) (n+3) K(f_1, h_1) K(f_2, h_2) \norm{ \psi^{(n)}(g) } \eqend{.}
\end{splitequation}
Iterating, it follows that
\begin{splitequation}
\label{eq:fixedparticlebound_k}
&\norm{ \prod_{j=1}^k \left[ \Phi(f_j) + \normord{ \Phi^2 }(h_j) \right] \psi^{(n)}(g) } \\
&\quad\leq 5^k \prod_{j=1}^k (n-1+2j) K(f_j, h_j) \norm{ \psi^{(n)}(g) } \\
&\quad= 10^k \frac{\Gamma\left( k + \frac{n+1}{2} \right)}{\Gamma\left( \frac{n+1}{2} \right)} \prod_{j=1}^k K(f_j, h_j) \norm{ \psi^{(n)}(g) } \eqend{.}
\end{splitequation}
Note also that by the definition~\eqref{eq:kalphabeta_def} of $K$ and the one of $c(h)$~\eqref{eq:ch_def}, we have
\begin{equation}
\label{eq:k_scaling}
K(\alpha f, \beta h) \leq \max(|\alpha|,|\beta|) K(f,h) \eqend{.}
\end{equation}

\section{Bounds on integral kernels}
\label{sec:bounds_kernels}

For a function $f(\vec{p},\vec{q})$ of two momenta we consider the seminorm
\begin{equation}
\label{eq:bound_kernel_norm}
\norm{ f }_{m,n} \defby \sup_{\substack{\abs{\alpha_1} + \abs{\alpha_2} \leq m \\ \abs{\beta_1} + \abs{\beta_2} \leq n}} \iint \abs{ p^{\alpha_1} q^{\alpha_2} \partial_\vec{p}^{\beta_1} \partial_\vec{q}^{\beta_2} f(\vec{p},\vec{q}) } \total \mu(\vec{p}) \total \mu(\vec{q}) \eqend{,}
\end{equation}
where we take on-shell momenta $p = (\omega_\vec{p},\vec{p})$ and $q = (\omega_\vec{q},\vec{q})$ and the measure $\total \mu(\vec{p})$ from \eqref{eq:hilbert_1p_measure}. Here, $\alpha_1, \alpha_2$ are $(D+1)$-dimensional multiindices and $\beta_1, \beta_2$ are $D$-dimensional multiindices. In the remainder of this section, we set the mass to $m = 1$.

\begin{lemma}
\label{lemma:bound_kernel}
For $N \in \mathbb{N}$, $\sigma = (\sigma_1,\ldots,\sigma_{N+1}) \in \{+1,-1\}^{N+1}$ and $h \in \mathcal{S}(\mathbb{R}^d)$ let
\begin{equation}
a_\sigma^{(N+1)}(\vec{p}_1,\vec{p}_{N+1}) \defby \int\cdots\int \prod_{j=1}^N \tilde{h}(\sigma_j p_j - \sigma_{j+1} p_{j+1}) \Bigr\rvert_{p_j^0 = \omega_{\vec{p}_j}} \prod_{j=2}^N \total \mu(\vec{p}_j) \eqend{.}
\end{equation}
If $\sigma$ is mixed, i.e., $\sigma \neq (+1,\ldots,+1)$ and $\sigma \neq (-1,\ldots,-1)$, there exist constants $C > 0$ and $k, \ell \in \mathbb{N}_0$ depending only on $m$ and $n$ such that
\begin{equation}
\norm{ a_\sigma^{(N+1)} }_{m,n} \leq \left[ C \norm{ h }_{\mathbb{R}^d,k,\ell} \right]^N \eqend{,}
\end{equation}
where $\norm{ h }_{\mathbb{R}^d,k,\ell}$ is the Schwartz seminorm defined in Eq.~\eqref{eq:schwartz_seminorm}.
\end{lemma}
\begin{proof}
We proceed by induction in $N$. For $N = 1$, since $\sigma$ is mixed we have
\begin{splitequation}
a_\sigma^{(2)}(\vec{p},\vec{q}) &= \tilde{h}\left( \pm (p+q) \right) \Bigr\rvert_{p^0 = \omega_\vec{p}, q^0 = \omega_\vec{q}} \\
&= \int h(x) \, \mathe^{\pm \mathi (\omega_\vec{p}+\omega_\vec{q}) t \mp \mathi (\vec{p}+\vec{q}) \vec{x}} \total^d x \eqend{.}
\end{splitequation}
Derivatives with respect to $\vec{p}$ or $\vec{q}$ acting on $a_\sigma^{(2)}(\vec{p},\vec{q})$ result in powers of $t$ and $\vec{x}$. Using then that
\begin{equation}
\mathe^{\pm \mathi (\omega_\vec{p}+\omega_\vec{q}) t} = \mp \frac{\mathi}{\omega_\vec{p} + \omega_\vec{q}} \frac{\partial}{\partial t} \mathe^{\pm \mathi (\omega_\vec{p}+\omega_\vec{q}) t} \eqend{,}
\end{equation}
we may introduce inverse factors of $\omega_\vec{p}+\omega_\vec{q}$ at the expense of derivatives with respect to $t$, which we integrate by parts. It follows that there exists constants $C > 0$ and $k,\ell \in \mathbb{N}_0$ such that
\begin{splitequation}
\abs{ \partial_\vec{p}^{\beta_1} \partial_\vec{q}^{\beta_2} a_\sigma^{(2)}(\vec{p},\vec{q}) } \leq C (\omega_\vec{p}+\omega_\vec{q})^{-j} \norm{ h }_{\mathbb{R}^d,k,\ell}
\end{splitequation}
for any $j \in \mathbb{N}_0$, which only depend on $\beta_1$, $\beta_2$ and $j$. Choosing $j = m + 2 d$ and replacing $C$, $k$ and $l$ by their suprema over all $\abs{\beta_1} + \abs{\beta_2} \leq n$, we thus obtain
\begin{splitequation}
\norm{ a_\sigma^{(2)} }_{m,n} &= \sup_{\substack{\abs{\alpha_1} + \abs{\alpha_2} \leq m \\ \abs{\beta_1} + \abs{\beta_2} \leq n}} \iint \abs{ p^{\alpha_1} q^{\alpha_2} \partial_\vec{p}^{\beta_1} \partial_\vec{q}^{\beta_2} a_\sigma^{(2)}(\vec{p},\vec{q}) } \total \mu(\vec{p}) \total \mu(\vec{q}) \\
&\leq C \sup_{\substack{\abs{\alpha_1} + \abs{\alpha_2} \leq m \\ \abs{\beta_1} + \abs{\beta_2} \leq n}} \iint \abs{ p^{\alpha_1} q^{\alpha_2} (\omega_\vec{p}+\omega_\vec{q})^{-m - 2 d} } \total \mu(\vec{p}) \total \mu(\vec{q}) \, \norm{ h }_{\mathbb{R}^d,k,\ell} \\
&\leq C \sup_{\substack{\abs{\alpha_1} + \abs{\alpha_2} \leq m \\ \abs{\beta_1} + \abs{\beta_2} \leq n}} \iint (\omega_\vec{p}+\omega_\vec{q})^{- 2 d} \total \mu(\vec{p}) \total \mu(\vec{q}) \, \norm{ h }_{\mathbb{R}^d,k,\ell} \\
&\leq C' \norm{ h }_{\mathbb{R}^d,k,\ell} \eqend{,}
\end{splitequation}
with a constant $C' > 0$ depending only on $m$ and $n$.

For the induction step, we note that at least one of $\sigma_> \defby (\sigma_2,\ldots,\sigma_{N+1})$ or $\sigma_< \defby (\sigma_1,\ldots,\sigma_N)$ is mixed. Assume that $\sigma_>$ is mixed, and consider the expansion
\begin{equation}
a_\sigma^{(N+1)}(\vec{p}_1,\vec{p}_{N+1}) = \int \tilde{h}(\sigma_1 p_1 - \sigma_2 p_2) \Bigr\rvert_{p_1^0 = \omega_{\vec{p}_1}, p_2^0 = \omega_{\vec{p}_2}} a_{\sigma_>}^{(N)}(\vec{p}_2,\vec{p}_{N+1}) \total \mu(\vec{p}_2) \eqend{.}
\end{equation}
Similarly to the $N = 1$ case, there exist constants $C' > 0$ and $k', \ell' \in \mathbb{N}_0$ such that
\begin{equation}
\abs{ \partial_{\vec{p}_1}^{\beta_1} \tilde{h}(\sigma_1 p_1 - \sigma_2 p_2) \Bigr\rvert_{p_1^0 = \omega_{\vec{p}_1}, p_2^0 = \omega_{\vec{p}_2}} } \leq C' \norm{h}_{\mathbb{R}^d,k',\ell'} \left( 1 + \abs{ \sigma_1 \vec{p}_1 - \sigma_2 \vec{p}_2 } \right)^{-j'}
\end{equation}
for any $j' \in \mathbb{N}_0$, where the constants depend on $\beta_1$ and $j'$. Using that
\begin{splitequation}
\abs{ p_1^{\alpha_1} } &\leq \omega_{\vec{p}_1}^{\abs{\alpha_1}} = \left( \vec{p}_1^2 + 1 \right)^\frac{\abs{\alpha_1}}{2} \leq \left( 2 \abs{ \sigma_1 \vec{p}_1 - \sigma_2 \vec{p}_2 }^2 + 2 ( \vec{p}_2 )^2 + 1 \right)^\frac{\abs{\alpha_1}}{2} \\
&\leq C'' \omega_{\vec{p}_2}^{\abs{\alpha_1}} \left( 1 + \abs{ \sigma_1 \vec{p}_1 - \sigma_2 \vec{p}_2 } \right)^{\abs{\alpha_1}}
\end{splitequation}
for a constant $C'' > 0$ depending on $\alpha_1$ and choosing $j' = \abs{\alpha_1} + d$, we obtain
\begin{splitequation}
&\int \abs{ p_1^{\alpha_1} \partial_{\vec{p}_1}^{\beta_1} \tilde{h}(\sigma_1 p_1 - \sigma_2 p_2) \Bigr\rvert_{p_1^0 = \omega_{\vec{p}_1}, p_2^0 = \omega_{\vec{p}_2}} } \total \mu(\vec{p}_1) \\
&\quad\leq C' C'' \omega_{\vec{p}_2}^{\abs{\alpha_1}} \norm{h}_{\mathbb{R}^d,k',\ell'} \int \left( 1 + \abs{ \sigma_1 \vec{p}_1 - \sigma_2 \vec{p}_2} \right)^{-d} \total \mu(\vec{p}_1) \\
&\quad\leq \frac{C' C''}{2 (2 \pi)^D} \, \omega_{\vec{p}_2}^{\abs{\alpha_1}} \norm{h}_{\mathbb{R}^d,k',\ell'} \int \left( 1 + \abs{ \vec{q} } \right)^{-d} \total^D \vec{q} \\
&\quad\leq C''' \omega_{\vec{p}_2}^{\abs{\alpha_1}} \norm{h}_{\mathbb{R}^d,k',\ell'} \eqend{,}
\end{splitequation}
where we performed the change of integration variable $\vec{p}_1 \to \vec{q} = \sigma_1 \vec{p}_1 - \sigma_2 \vec{p}_2$. Here, the constants $C'''$, $k'$ and $\ell'$ depend only on $\alpha_1$ and $\beta_1$, and replacing them by their maximum over all $\abs{ \alpha_1 } \leq m$, $\abs{ \beta_1 } \leq n$ and using dominated convergence, it follows that
\begin{splitequation}
&\norm{ a_\sigma^{(N+1)} }_{m,n} \\
&\quad\leq C''' \norm{h}_{\mathbb{R}^d,k',\ell'} \sup_{\substack{\abs{\alpha_1} + \abs{\alpha_2} \leq m \\ \abs{\beta_2} \leq n}} \iint \abs{ \omega_{\vec{p}_2}^{\abs{\alpha_1}} p_{N+1}^{\alpha_2} \partial_{\vec{p}_{N+1}}^{\beta_2} a_{\sigma_>}^{(N)}(\vec{p}_2,\vec{p}_{N+1}) } \total \mu(\vec{p}_2) \total \mu(\vec{p}_{N+1}) \\
&\quad\leq C''' \norm{h}_{\mathbb{R}^d,k',\ell'} \norm{ a_{\sigma_>}^{(N)} }_{m,n} \eqend{.}
\end{splitequation}
By the induction hypothesis, it holds that $\norm{ a_{\sigma_>}^{(N)} }_{m,n} \leq \left[ C \norm{ h }_{\mathbb{R}^d,k,\ell} \right]^{N-1}$ with constants $C$, $k$ and $\ell$ only depending on $m$ and $n$, and we therefore obtain
\begin{equation}
\norm{ a_\sigma^{(N+1)} }_{m,n} \leq \left[ \max(C,C''') \norm{ h }_{\mathbb{R}^d,\max(k,k'),\max(\ell,\ell')} \right]^N
\end{equation}
as required. The analogous steps apply if $\sigma_<$ is mixed, and we obtain the conclusion.
\end{proof}

\section{Rearrangements of absolutely convergent series}
\label{app:pringsheim}

It is well known that for any series of real numbers which does not converge absolutely, there exists a rearrangement such that the new series converges to any given value, or diverges (the Riemann rearrangement theorem). For double series, similar results hold, which were first given by Pringsheim~\cite{pringsheim1899}. In particular, given a double series $\sum_{k,l=0}^\infty a_{kl}$ he shows that one can perform the sum in any order if it converges absolutely. In this section, we show that other rearrangements are also possible, in particular diagonal ones, and we generalize to higher-order series.

\begin{lemma}
\label{lem:pringsheim}
Let $A = \sum_{k_1,k_2,\ldots,k_n=0}^\infty a_{k_1 \cdots k_n}$ be absolutely convergent. Consider a rearrangement $R$ of summation indices, i. e., a sequence $R = ( R(r) )_{r=0}^\infty$ of finite sets of $n$-tuples $\vec{k} = (k_1,\ldots,k_n)$ which is such that $R(r) \cap R(r') = \emptyset$ for $r \neq r'$, and that for each $\vec{k}$ there is some $r \in \mathbb{N}$ with $\vec{k} \in R(r)$. Then the rearrangement
\begin{equation}
\sum_{r=0}^\infty \sum_{\vec{k} \in R(r)} a_{\vec{k}} = A
\end{equation}
of the series converges to the same value.
\end{lemma}
The second condition on the rearrangement ensures that eventually all cubes are summed, i.e., that for each $s \in \mathbb{N}$ there exists $r = r(s) \in \mathbb{N}$ such that
\begin{equation}
\label{eq:pringsheim_cubes}
\{ 0,\ldots,s \}^{\times n} \subseteq \bigcup_{r'=1}^{r(s)} R(r') \eqend{.}
\end{equation}
For example, the diagonal rearrangement $R(r) = \{ \vec{k} \colon \abs{\vec{k}} \defby k_1 + \cdots + k_n = r \}$ fulfills this with $r(s) = n s$.
\begin{proof}
We set
\begin{equation}
A_{m_1 \cdots m_n} \defby \sum_{k_1=0}^{m_1} \cdots \sum_{k_n=0}^{m_n} a_{k_1 \cdots k_n} \eqend{,} \quad \abs{A}_{m_1 \cdots m_n} \defby \sum_{k_1=0}^{m_1} \cdots \sum_{k_n=0}^{m_n} \abs{ a_{k_1 \cdots k_n} } \eqend{.}
\end{equation}
The given sum is convergent with value $A$ if for each $\epsilon > 0$ there exist $N_1,\ldots,N_n \in \mathbb{N}$ such that
\begin{equation}
\abs{ A_{m_1 \cdots m_n} - A } < \epsilon \quad\text{for all}\quad m_1 \geq N_1, \ldots, m_n \geq N_n \eqend{.}
\end{equation}
Clearly we can also replace $< \epsilon$ by $\leq \epsilon$ and the condition by $m_1 \geq N, \ldots, m_n \geq N$ for a single $N \in \mathbb{N}$. Since by assumption the sum is absolutely convergent (with value $B$, say), the same holds for $\abs{A}_{m_1 \cdots m_n}$: for each $\epsilon > 0$ there exists an $N \in \mathbb{N}$ such that
\begin{equation}
\abs{ \abs{A}_{m_1 \cdots m_n} - B } \leq \epsilon \quad\text{for all}\quad m_1 \geq N, \ldots, m_n \geq N \eqend{.}
\end{equation}

Since
\begin{equation}
\abs{ \sum_{r'=0}^r \sum_{\vec{k} \in R(r')} a_{\vec{k}} } \leq \sum_{r'=0}^r \sum_{\vec{k} \in R(r')} \abs{a_{\vec{k}}} \eqend{,}
\end{equation}
we first show that the right-hand side converges to $B$ as $r \to \infty$. Given $\epsilon > 0$, choose $N$ such that $\abs{ \abs{A}_{N \cdots N} - B } \leq \epsilon$, and $r = r(N)$ such that~\eqref{eq:pringsheim_cubes} holds. Since $R(r)$ is a finite set, there also exists $M = M(r)$ such that
\begin{equation}
\bigcup_{r'=1}^r R(r') \subseteq \{ 0,\ldots,M(r) \}^{\times n} \eqend{,}
\end{equation}
and we obtain
\begin{splitequation}
\abs{ \sum_{r'=0}^{r(N)} \sum_{\vec{k} \in R(r')} \abs{a_{\vec{k}}} - B } &\leq \sum_{r'=0}^{r(N)} \sum_{\vec{k} \in R(r')} \abs{a_{\vec{k}}} - \abs{A}_{N \cdots N} + \abs{ \abs{A}_{N \cdots N} - B } \\
&\leq \abs{A}_{M(r(N)) \cdots M(r(N))} - \abs{A}_{N \cdots N} + \abs{ \abs{A}_{N \cdots N} - B } \\
&\leq \abs{ \abs{A}_{M(r(N)) \cdots M(r(N))} - B } + 2 \abs{ \abs{A}_{N \cdots N} - B } \leq 3 \epsilon \eqend{,}
\end{splitequation}
since obviously $M(r(N)) \geq N$. Since $\epsilon$ was arbitrary, we obtain
\begin{equation}
\lim_{r \to \infty} \sum_{r'=0}^r \sum_{\vec{k} \in R(r')} \abs{a_{\vec{k}}} = B = \sum_{k_1,k_2,\ldots,k_n=0}^\infty \abs{a_{k_1 \cdots k_n}} \eqend{.}
\end{equation}

The analogous argument then shows that the series itself is convergent with the same sum:
\begin{splitequation}
\abs{ \sum_{r'=0}^{r(N)} \sum_{\vec{k} \in R(r')} a_{\vec{k}} - A } &\leq \abs{ \sum_{r'=0}^{r(N)} \sum_{\vec{k} \in R(r')} a_{\vec{k}} - A_{N \cdots N} } + \abs{ A_{N \cdots N} - A } \\
&= \abs{ \sum_{\vec{k} \in \bigcup_{r'=0}^{r(N)} R(r') \setminus \{ 0,\ldots,N \}^{\times n}} a_\vec{k} } + \abs{ A_{N \cdots N} - A } \\
&\leq \sum_{\vec{k} \in \cup_{r'=0}^{r(N)} R(r') \setminus \{ 0,\ldots,N \}^{\times n}} \abs{ a_\vec{k} } + \abs{ A_{N \cdots N} - A } \\
&\leq \abs{A}_{M(r(N)) \cdots M(r(N))} - \abs{A}_{N \cdots N} + \abs{ A_{N \cdots N} - A } \\
&\leq \abs{ \abs{A}_{M(r(N)) \cdots M(r(N))} - B } + \abs{ \abs{A}_{N \cdots N} - B } + \abs{ A_{N \cdots N} - A } \leq 3 \epsilon \eqend{,}
\end{splitequation}
where $N$ is such that both $\abs{ A_{n \cdots n} - A } \leq \epsilon$ and $\abs{ \abs{A}_{n \cdots n} - B } \leq \epsilon$ for all $n \geq N$. Therefore, we also have
\begin{equation}
\lim_{r \to \infty} \sum_{r'=0}^r \sum_{\vec{k} \in R(r')} a_{\vec{k}} = A = \sum_{k_1,k_2,\ldots,k_n=0}^\infty a_{k_1 \cdots k_n} \eqend{.}
\end{equation}
\end{proof}

\bibliography{arxiv_submit_v2}

\end{document}